\documentclass[prc,aps,twocolumn,superscriptaddress,showpacs,amsmath,amssymb,nofootinbib]{revtex4-2}
\usepackage[utf8]{inputenc}
\usepackage{graphicx}
\usepackage[breaklinks, colorlinks=true]{hyperref}
\hypersetup{
 linkcolor=blue, citecolor=cyan}
\usepackage{bm}
\usepackage{siunitx}
\usepackage{mathtools}
\usepackage[normalem]{ulem}

\newcommand{\Tf}{T_\mathrm{f}}
\newcommand{\tauf}{\tau_\mathrm{f}}

\begin{document}



\title{Local Spin Polarization in Anisotropic Gubser Flow:\\
Suppression Mechanism and Formulation Dependence}

\author{Kenji Fukushima}
\email{fuku@nt.phys.s.u-tokyo.ac.jp}
\affiliation{Department of Physics, The University of Tokyo, 
  7-3-1 Hongo, Bunkyo-ku, Tokyo 113-0033, Japan}

\author{Shi Pu}
\email{shipu@ustc.edu.cn}
\affiliation{Department of Modern Physics, University of Science and Technology of China, Anhui 230026, China}
\affiliation{Southern Center for Nuclear-Science Theory (SCNT), Institute of Modern Physics, Chinese Academy of Sciences, Huizhou 516000, Guangdong Province,
China}

\author{Dong-Lin Wang}
\email{donglinwang@mail.ustc.edu.cn}
\affiliation{Department of Modern Physics, University of Science and Technology of China, Anhui 230026, China}
\affiliation{Department of Physics, The University of Tokyo, 
  7-3-1 Hongo, Bunkyo-ku, Tokyo 113-0033, Japan}

\begin{abstract}
\end{abstract}

\begin{abstract}
  We analytically study the longitudinal spin polarization in relativistic heavy-ion collisions using a perturbed Gubser flow solution.  In the large-system-size limit, we derive anaytical expression of the local spin polarization along the beam direction.
  In our treatment, the contributions from thermal vorticity and thermal shear are of comparable magnitudes.  The thermal vorticity yields the polarization with a sign opposite to that observed experimentally, while the thermal shear counteracts this effect, helping recover the desired sign.  We find that the choice of the reference unit vector aligned with the fluid velocity gives the experimentally observed sign only at low transverse momenta, whereas another formulation with the unit vector fixed along the laboratory time direction yields the desired sign for a wide range of transverse momenta.  Notably, a recent formulation with the unit vector normal to the freeze-out hypersurface exhibits an exact cancellation between contributions from thermal vorticity and thermal shear at leading order in the large-system-size limit.  We identify a general cancellation pattern with acceleration dominance, which is manifest particularly in the latter two formulations.  Thus, the total polarization originates from non-acceleration effects, which need not be substantial even when the elliptic flow is finite, as clearly demonstrated in our analytical results.  For comparison, we also discuss the spin polarization in the Hubble flow with rotation.
\end{abstract}
\maketitle

\section{Introduction}

In noncentral relativistic heavy-ion collisions, the quark-gluon plasma
(QGP) contains a large orbital angular momentum oriented perpendicular
to the reaction plane. Part of the orbital angular momentum is subsequently
transferred to the spin of produced particles, e.g., $\Lambda$ hyperons,
causing an average polarization along the direction of the orbital
angular momentum \cite{Liang:2004ph,Gao:2007bc,Voloshin:2004ha}.
This phenomenon is known as global spin polarization. Experimental
measurements from STAR Collaboration \cite{STAR:2017ckg}, ALICE
Collaboration \cite{ALICE:2019onw}, and HADES Collaboration \cite{HADES:2022enx}
have already confirmed the existence of the global spin polarization.
Theoretically, the experimental data can be well described by various
phenomenological models \cite{Karpenko:2016jyx,Li:2017slc,Sun:2017xhx,Wei:2018zfb,Vitiuk:2019rfv,Fu:2020oxj,Lei:2021mvp,Ryu:2021lnx,Wu:2022mkr}
using the modified Cooper-Frye formula with the thermal vorticity
\cite{Becattini:2013fla,Fang:2016vpj}, confirming the mutual conversion
between the orbital angular momentum of QGP and the spin of produced
particles. 

However, the successful theoretical description encounters challenges
when extended to local spin polarization, which is a function of the produced
particle's momentum. Along the beam direction, the local spin polarization
induced by elliptic flow and triangular flow has already been measured
\cite{STAR:2019erd,ALICE:2021pzu,STAR:2023eck,CMS:2025nqr}. Compared
to the experimental results, theoretical calculations based on thermal
vorticity alone give an opposite sign in the azimuthal-angle dependence
\cite{Becattini:2017gcx,Xia:2018tes,Florkowski:2019voj,Wu:2019eyi,Xia:2019fjf,Becattini:2019ntv}.
This discrepancy, called the \textit{sign puzzle}, implies that the original
method based on thermal vorticity alone is insufficient to describe
the local observable. The puzzle has stimulated substantial theoretical
activity, accelerating the development of spin physics in heavy-ion
collisions. For example, quantum kinetic theory
\cite{Gao:2012ix,Chen:2012ca,Hidaka:2016yjf,Hidaka:2017auj,Gao:2019znl,Weickgenannt:2019dks,Liu:2020flb,Weickgenannt:2020aaf,Weickgenannt:2021cuo,Sheng:2021kfc,Hidaka:2022dmn,Fang:2022ttm,Dong:2022yzt,Fang:2023bbw,Fang:2024vds}
and relativistic spin hydrodynamics \cite{Montenegro:2017lvf,Florkowski:2017ruc,Becattini:2018duy,Florkowski:2018fap,Hattori:2019lfp,Fukushima:2020ucl,Li:2020eon,Hongo:2021ona,She:2021lhe,Wang:2021ngp,Weickgenannt:2022qvh,Weickgenannt:2022zxs,Cao:2022aku,Bhadury:2022ulr,Florkowski:2024bfw,Huang:2024ffg,Fang:2025aig}
have witnessed rapid development in recent years. 

To address the sign puzzle, theoretical frameworks have been extended by
incorporating the thermal shear tensor into the modified Cooper-Frye
formula \cite{Liu:2021uhn,Becattini:2021suc,Hidaka:2017auj}. The subsequent numerical
simulations \cite{Fu:2021pok,Becattini:2021iol,Yi:2021ryh} reveal that the
polarization induced by thermal shear has a sign opposite to the polarization
from thermal vorticity, implying a cancellation when adding them together.
Under certain conditions, the thermal shear contribution exceeds that
of the thermal vorticity, thereby recovering the correct sign of the
local spin polarization \cite{Fu:2021pok,Becattini:2021iol,Yi:2021ryh,Ryu:2021lnx,Florkowski:2021xvy,Wu:2022mkr,Buzzegoli:2022fxu,Palermo:2022lvh,Palermo:2024tza,Yi:2024kwu}.
For a comprehensive overview, we refer readers to the review articles
\cite{Hidaka:2022dmn,Becattini:2024uha,Niida:2024ntm}.

Currently, several distinct formulations have been proposed to quantify
the thermal shear effect on local spin polarization. The main difference
among them stems from the choice of the unit vector in the thermal
shear term. Specifically, the Fu-Liu-Pang-Song-Yin (FLPSY) formulation
\cite{Liu:2021uhn,Fu:2021pok} aligns this vector with the fluid
velocity, whereas the Becattini-Buzzegoli-Palermo-Inghirami-Karpenko
(BBPIK) formulation \cite{Becattini:2021suc,Becattini:2021iol} sets
it to be the time direction in the laboratory frame. The Sheng-Becattini-Roselli
(SBR) formulation \cite{Sheng:2025cjk} defines the vector as the
normal to the freeze-out hypersurface. While these formulations share
the common goal, they lead to quantitatively different predictions.
A systematic comparison of these formulations is therefore warranted.
To date, numerical comparisons of the first two formulations have been
conducted in some cases \cite{Alzhrani:2022dpi,Yi:2024kwu}. However,
it is still challenging to uncover underlying mechanisms from the
numerical results. 

In contrast, analytical studies can explicitly show how the local spin
polarization scales with initial conditions and transverse momentum,
as well as the cancellation between thermal shear and thermal vorticity
effects. At present, the existing analytical studies addressing local
spin polarization are based on the modified blast-wave model \cite{Arslan:2024dwi,Arslan:2025tan},
in which the dynamical variables are decoupled from the hydrodynamic
equations. To the best of our knowledge, there are currently no analytic
studies of spin polarization derived directly from hydrodynamic solutions. 

In this work, we fill this gap by studying longitudinal spin polarization
using a perturbed analytical solution of relativistic hydrodynamics.
Specifically, we will use an anisotropic extension of the Gubser flow
that incorporates both elliptic and triangular geometric deformations
\cite{Gubser:2010ze,Gubser:2010ui,Hatta:2014jva,Ren:2026fqj}. We will derive
explicit analytical expressions for the spin polarization
for three formulations. This analytical approach enables us to systematically
examine the parameter dependence and perform a direct comparison among
these formulations. We aim to clarify the conditions under which the
desired sign of local spin polarization is recovered, and to identify
potential cancellation between thermal shear and thermal
vorticity effects.

The paper is organized as follows. In Sec.~\ref{sec:Analytical-Fluid-Solution},
we present several analytical hydrodynamic solutions. In Sec.~\ref{sec:Three-Different-Formulations},
we review the three formulations for spin polarization. In Sec.~\ref{sec:Spin-Polarization-Hubble}, we calculate the spin polarization in the rotating Hubble flow.
In Sec.~\ref{sec:Spin-Polarization-from}, we analyze the longitudinal spin polarization in the perturbed Gubser flow. Finally, we summarize our findings and discuss their implications in Sec.~\ref{sec:Conclusion}.
Throughout this work, we adpot the metric $g_{\mu\nu}=\mathrm{diag}\{+,-,-,- \}$ for flat space-time.

\section{Analytical Fluid Solutions}
\label{sec:Analytical-Fluid-Solution}

We focus on ideal relativistic hydrodynamics without charge
currents for simplicity.  We can generalize our approach to viscous hydrodynamics if needed.  In the ideal case, the energy-momentum tensor is decomposed as 
\begin{equation}
  T^{\mu\nu} = e u^{\mu}u^{\nu} - p\Delta^{\mu\nu} \,,
\end{equation}
where $u^{\mu}$, $e$, and $p$ represent the fluid velocity, the energy density, and the thermodynamic pressure, respectively.  In our convention the projection tensor is defined as $\Delta^{\mu\nu}=g^{\mu\nu}-u^{\mu}u^{\nu}$.  The energy-momentum tensor satisfies the
conservation law:
\begin{equation}
  \nabla_{\mu} T^{\mu\nu} = 0 \,,
  \label{eq:conservedEq0}
\end{equation}
and the equation of state is required to close the system of equations.  Here, we adopt the conformal equation of state, i.e.,
\begin{equation}
  e = 3p = \lambda T^{4} \,,
  \label{eq:Eos}
\end{equation}
where $\lambda$ is a dimensionless constant and $T$ is the fluid temperature.

\begin{figure}
  \centering
  \includegraphics[width=0.98\columnwidth]{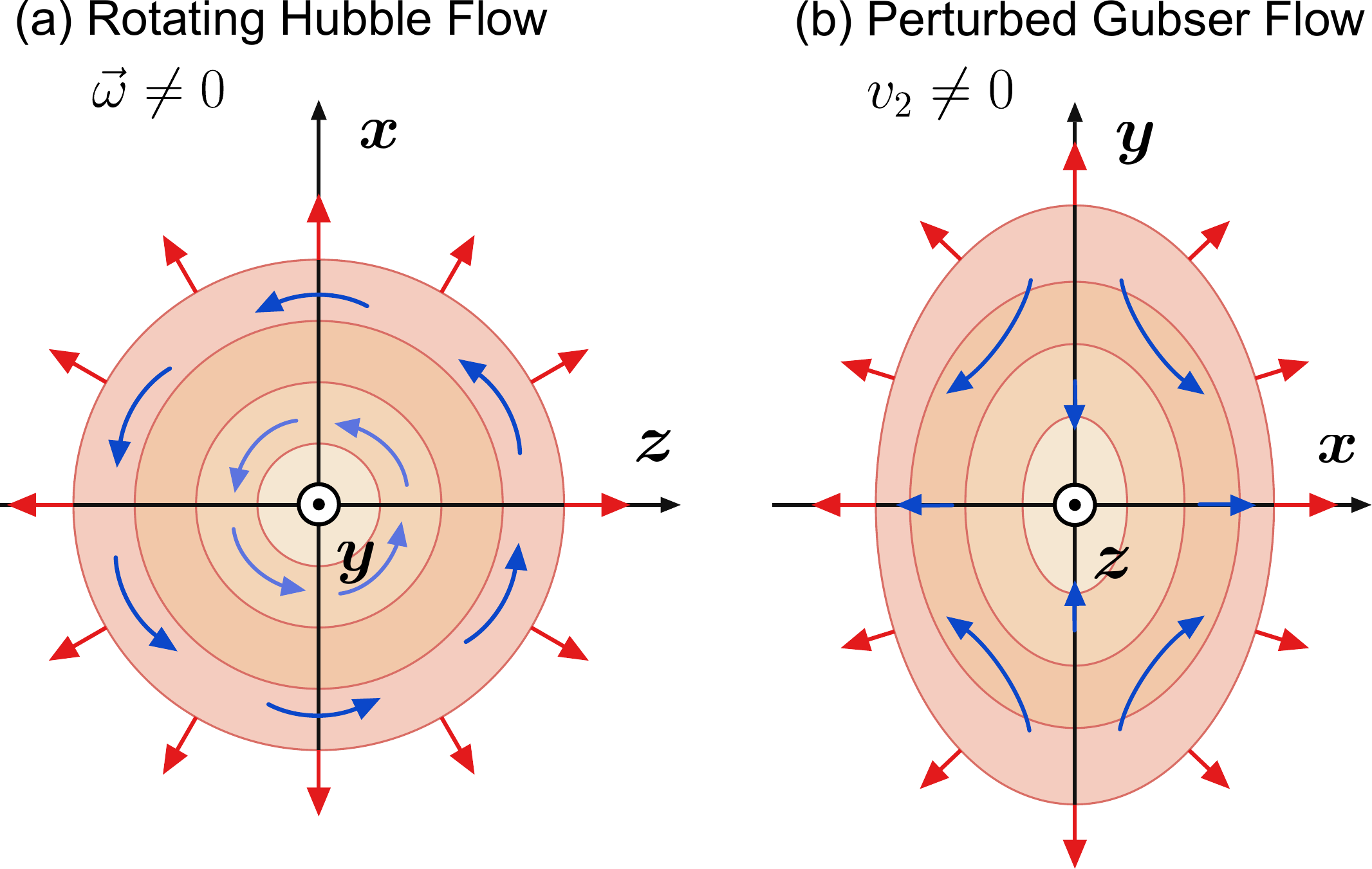}
  \caption{Schematic illustration of the rotating Hubble flow (left) and the perturbed Gubser flow (right). The rotating Hubble flow comprises spherical expansion (red arrows) with rigid rotation (blue arrows) about the $y$-axis. The perturbed Gubser flow builds upon the original Gubser flow (red arrows) by introducing additional anisotropy (blue arrows) in the $xy$-plane.}
  \label{fig:flows}
\end{figure}

We are constructing the analytical solutions of relativistic hydrodynamics to approximate the QGP in heavy-ion collisions.  The rotating Hubble flow and the perturbed Gubser flow are two representative examples, as schematically illustrated in Fig.~\ref{fig:flows}.  The former is a simple solution with spherical expansion and rigid rotation at finite angular velocity, $\vec{\omega}$, qualitatively corresponding to global polarization, while the latter is a solution of our present interest with longitudinal and transverse expansion, including a nonzero elliptic flow, $v_2$.  In the subsequent sections, we will explain these two solutions in order.

\subsection{Rotating Hubble Flow}

For spherically expanding systems, the Hubble flow is a simple solution of the relativistic hydrodynamic equations~\cite{Csorgo:2003rt}. This solution can be generalized to include rotation as~\cite{Hatta:2014gga}
\begin{equation}
  (u^{t}, \vec{u}) = \gamma_{\omega}(t, \vec{x}+\vec{\omega}\times\vec{x}) \,,\qquad
  T = T_{0} \gamma_{\omega} \,,
  \label{eq:HubbleFlowSol}
\end{equation}
where $\vec{\omega}$ is the angular velocity vector, $T_{0}$ is a dimensionless constant determined by the initial condition, and $\gamma_{\omega}=[t^{2}-(\vec{x}+\vec{\omega}\times\vec{x})^{2}]^{-1/2}$.  The velocity profile captures a spherically symmetric expansion coupled with rigid rotation along the axis parallel to $\vec{\omega}$.  In the $\vec{\omega}\to0$ limit, Eq.~\eqref{eq:HubbleFlowSol} reduces to the standard Hubble flow.  This exact solution provides an analytical toy model for relativistic fluids with rigid rotation and can be applied, for example, in the cosmological context~\cite{Weinberg:1972kfs}.  Also in the heavy-ion collision, the Hubble flow is a qualitatively reasonable description as long as the collision energy is low, while higher energy collisions would produce a fireball expanding predominantly along the beam direction with approximate boost invariance.

\subsection{Bjorken Flow}

For boost-invariant and cylindrically symmetric systems,  the Bjorken flow~\cite{Bjorken:1982qr} is the best-known.  To describe the Bjorken flow, it is natural to work in Milne
coordinates $(\tau,r,\phi,\eta)$,
where $\tau=\sqrt{t^{2}-z^{2}}$ is the longitudinal proper time,
$r=\sqrt{x^{2}+y^{2}}$ is the transverse radius, $\phi=\tan^{-1}(y/x)$
is the azimuthal angle, and $\eta=\tanh^{-1}(z/t)$ is the rapidity.  The corresponding metric reads:
\begin{equation}
  ds^{2} = d\tau^{2} - dr^{2} - r^{2} d\phi^{2} - \tau^{2} d\eta^{2} \,.
  \label{eq:Metric0}
\end{equation}
In Milne coordinates, the velocity profile of the Bjorken flow takes a
simple form~\cite{Bjorken:1982qr}: 
\begin{equation}
  u^{\mu} = (1,\boldsymbol{0}) \,,\qquad T=T_{0}\tau^{-1} \,.
\end{equation}
Although the Bjorken flow is useful to approximate bulk properties of the hydrodynamic evolutions at high enough energy, it is too simple to discuss spin polarization, as the $(1+1)$-dimensional solution inevitably yields zero spin polarization.  To obtain a nontrivial result, we should consider more realistic hydrodynamic solutions.
Since transverse expansion is crucial for local spin polarization phenomena, we need to include the transverse dynamics, which has been implemented in the Gubser flow~\cite{Gubser:2010ze,Gubser:2010ui}.  For our purpose, moreover, it is also indispensable to introduce transverse anisotropy, which has been discussed perturbatively~\cite{Hatta:2014jva}.  We will explain constructions of these extensions in order.

\subsection{Gubser Flow}

The Gubser flow is a conformal solution with longitudinal and transverse expansion.  It is convenient to formulate it in terms of $\hat{x}^\mu$ on $dS_{3}\times\mathbb{R}$.
The Minkowski spacetime, $\mathbb{R}^{3,1}$,
is transformed to $dS_{3}\times\mathbb{R}$ by a Weyl rescaling, so that the line element in Eq.~\eqref{eq:Metric0} is deformed as
\begin{equation}
  d\hat{s}^{2} = \tau^{-2}ds^{2} = d\rho^{2}-\cosh^{2}\rho(d\theta^{2}+\sin^{2}\theta d\phi^{2})-d\eta^{2} \,,
  \label{eq:metricDs}
\end{equation}
where $\rho$ and $\theta$ are defined via
\begin{equation}
  \sinh\rho = -\frac{L^{2}-\tau^{2}+r^{2}}{2L\tau} \,,\quad
  \tan\theta = \frac{2Lr}{L^{2}+\tau^{2}-r^{2}} \,.
\end{equation}
We will explain the meaning of the parameter $L$ later.  With this Weyl rescaling, the temperature and fluid velocity transform as 
\begin{equation}
  \hat{T} = \tau T \,,
  \qquad
  \hat{u}_{\mu} = \frac{1}{\tau}\frac{\partial x^{\nu}}{\partial\hat{x}^{\mu}}u_{\nu} \,,
  \label{eq:transformationTu}
\end{equation}
where $\hat{x}^{\mu}=(\rho,\theta,\phi,\eta)$ and $x^{\mu}=(\tau,r,\phi,\eta)$.
The conservation equation~\eqref{eq:conservedEq0} with the
equation of state~\eqref{eq:Eos} leads to
\begin{subequations}
\begin{align}
  & \hat{T}\hat{u}^{\mu}\hat{\nabla}_{\mu}\hat{u}^{\alpha}-\hat{\Delta}^{\alpha\mu}\hat{\nabla}_{\mu}\hat{T} = 0 \,,
  \label{eq:eom01} \\
  & \hat{u}^{\mu}\hat{\nabla}_{\mu}\hat{T}+\frac{1}{3}\hat{T}\hat{\nabla}_{\mu}\hat{u}^{\mu} = 0 \,.
  \label{eq:eom02}
\end{align}
  \label{eq:eom}
\end{subequations}
The projection tensor acting on $\hat{x}^\mu$ is $\hat{\Delta}^{\mu\nu}=\hat{g}^{\mu\nu}-\hat{u}^{\mu}\hat{u}^{\nu}$ with the metric $\hat{g}^{\mu\nu}$ deduced from Eq.~\eqref{eq:metricDs}.  We note that $\hat{\nabla}_{\mu}$ is the covariant derivative defined in terms of $\hat{g}^{\mu\nu}$.

The fluid velocity profile in the Gubser flow is given in the $\hat{x}^\mu$ coordinates by
\begin{equation}
  (\hat{u}_{\rho}, \hat{u}_{\theta}, \hat{u}_{\phi}, \hat{u}_{\eta}) = (1,0,0,0) \,.
  \label{eq:GubserVelocity0}
\end{equation}
This solution trivially satisfies Eq.~\eqref{eq:eom01}, while Eq.~\eqref{eq:eom02} results in
\begin{equation}
  \hat{T} = \hat{T}_{0}(\cosh\rho)^{-2/3} \,,
  \label{eq:GubserT}
\end{equation}
where $\hat{T}_{0}$ is determined by the initial condition.

We can grasp an intuitive picture by transforming the solution back to Minkowski spacetime using Eq.~\eqref{eq:transformationTu}.  
The profile in the Milne coordinates becomes 
\begin{equation}
  (u_{\tau}, u_{r}, u_{\phi}, u_{\eta}) = \left( \frac{L^{2}+\tau^{2}+r^{2}}{2L\tau\cosh\rho}, \frac{-r}{L\cosh\rho}, 0, 0 \right) \,,
  \label{eq:Gubseru}
\end{equation}
from which we see that the fluid velocity has both longitudinal and transverse radial expansion.  It is easy to confirm that the large-$L$ limit reduces the Gubser flow to the Bjorken flow as $\cosh\rho \to L/(2\tau)$ for $L\to\infty$.  It should be emphasized that the expansion of the 
Gubser flow is still isotropic on the transverse plane.

The above expression of $u_r$ implies that $L$ is a typical scale along the $r$ direction.  Indeed, the rescaled energy density behaves as $\hat{e} \propto (\cosh\rho)^{-8/3}$. 
In the early time regime $\tau\approx0$, the energy density $\hat{e}$ decays along the $r$ direction as
\begin{equation}
  \frac{\hat{e}(r=L)}{\hat{e}(r=0)} \approx 0.16 \,.
\end{equation}
Thus, the parameter $L$ characterizes the initial transverse size of the QGP, inside of which the energy density is localized.

\subsection{Perturbed Gubser Flow}

We need to break transverse isotropy to describe local spin polarization phenomena.  For this purpose, we construct an approximate solution with perturbative anisotropy on top of the Gubser flow.  Following Refs.~\cite{Gubser:2010ui,Hatta:2014jva},
we consider the perturbed solution parametrized as
\begin{equation}
  (\hat{u}_{\rho}, \hat{u}_{\theta}, \hat{u}_{\phi}, \hat{u}_{\eta})
  = \bigl(1, v_{\theta}(\rho,\theta,\phi), v_{\phi}(\rho,\theta,\phi), 0\bigr),
\end{equation}
with $|v_{\theta,\phi}|\ll 1$ for the fluid velocity, and the local temperature is modified accordingly as
\begin{equation}
  \hat{T}(\rho,\theta,\phi) = \hat{T}_{0} (\cosh\rho)^{-2/3} \bigl[1+\mathcal{T}(\rho,\theta,\phi) \bigr],
\end{equation}
with $|\mathcal{T}|\ll 1$.  Substituting them into the conservation equations~\eqref{eq:eom}, we obtain:
\begin{subequations}
\begin{align}
  & 3\partial_{\rho} v_{\theta} - 2v_{\theta}\tanh\rho - 3\partial_{\theta}\mathcal{T} = 0\,,
  \label{eq:perturbEq01}\\
  & 3\partial_{\rho} v_{\phi} - 2v_{\phi}\tanh\rho - 3\partial_{\phi}\mathcal{T} = 0\,,
  \label{eq:perturbEq02}\\
  & 3\partial_{\rho}\mathcal{T} - \mathrm{sech}^{2}\rho(\partial_{\theta}v_{\theta} + v_{\theta}\cot\theta+\partial_{\phi}v_{\phi}\csc^{2}\theta) = 0
  \label{eq:perturbEq03}
\end{align}
  \label{eq:perturbEq}
\end{subequations}
up to the linear order in the perturbative expansion.

We can analytically solve the linear differential equations~\eqref{eq:perturbEq} using the method of separation of variables.  The general solution takes the following form: 
\begin{subequations}
\begin{align}
  v_{\theta} &= \sum_{l=0}^{\infty} \chi^{(l)}(\rho) \, \partial_{\theta}X^{(l)}(\theta,\phi) \,,
  \label{eq:Solvtheta}\\
  v_{\phi} &= \sum_{l=0}^{\infty} \chi^{(l)}(\rho) \, \partial_{\phi}X^{(l)}(\theta,\phi) \,,
  \label{eq:Solvphi}\\
  \mathcal{T} &= \sum_{l=0}^{\infty} \sigma^{(l)}(\rho)X^{(l)}(\theta,\phi) \,,
  \label{eq:SolT}
\end{align}
\label{eq:sol}
\end{subequations}
where 
\begin{equation}
  \begin{split}
    & X^{(l)}(\theta,\phi) \\
    & =\sum_{m=0}^{l} P_{l}^{m}(\cos\theta) \bigl[ c_{lm}\cos(m\phi) + d_{lm}\sin(m\phi) \bigr] 
    \label{eq:SolS}
  \end{split}
\end{equation}
with constant coefficients $c_{lm}$ and $d_{lm}$.  As usual, $P_{l}^{m}(x)$
represents the associated Legendre polynomial.

The functions $\chi^{(l)}$
and $\sigma^{(l)}$ satisfy, respectively:
\begin{align}
  & 3\frac{d}{d\rho}\chi^{(l)}-2\chi^{(l)}\tanh\rho-3\sigma^{(l)} = 0 \,,
  \label{eq:eqchi}\\
  & 3\frac{d}{d\rho}\sigma^{(l)}+l(l+1)\chi^{(l)}\mathrm{sech}^{2}\rho = 0 \,.
  \label{eq:eqsigma}
\end{align}
It would be technically difficult to cope with the analytical solutions of Eqs.~\eqref{eq:eqchi} and \eqref{eq:eqsigma}. Instead, to simplify them, let us focus on the large-size case \cite{Hatta:2014jva}, i.e.,
\begin{equation}
  \tau/L \ll 1 \,,
  \label{eq:limitc}
\end{equation}
which corresponds to $\rho\rightarrow-\infty$.  In this case, the leading-order asymptotic solution of Eqs.~\eqref{eq:eqchi} and \eqref{eq:eqsigma} is  
\begin{equation}
  \chi^{(l)} \simeq \frac{3}{2}\,,
  \qquad
  \sigma^{(l)}(\rho) \simeq 1\,,
  \label{eq:approxForChiSigma}
\end{equation}
where an overall multiplicative factor has been absorbed into the
coefficients $c_{lm}$ and $d_{lm}$. 

The choice of $c_{lm}$ and $d_{lm}$ depends on the physical setup of our interest.  For our present purpose, we will focus on a special choice with $m=l$, leading to
\begin{equation}
  X^{(l)}(\theta,\phi) = c_{ll} P_{l}^{l}(\cos\theta)\cos(l\phi)
  \label{eq:SolS1}
\end{equation}
for $l=2,3$ only.  This choice is motivated by the heavy-ion collision geometry.
For non-central collisions, the nuclear overlapping region in
the transverse plane is approximately elliptical~\cite{Yagi:2005yb}.  Without loss of generality, we can take the shorter side along the $x$ axis (i.e., $\phi=0,\pi$) and the longer side along the $y$ axis (i.e., $\phi=\pi/2,3\pi/2$).  To construct a solution with elliptic deformation, the $\phi$ dependence is dominated by terms with $m=2$ in Eq.~\eqref{eq:SolS}, among which $\sin(2\phi)$ should be removed because they violate the reflection
symmetry along the $x$ and $y$ axes; thus, $d_{l2}=0$ is fixed.

We conclude that $l=2$ should give the dominant contribution as follows.  Physically, the isothermal geometry is deformed elliptically, and for a fixed value of $r>0$, the temperature on the $x$ axis should be lower than that on the $y$ axis, namely,
\begin{equation}
  \mathcal{T}\,|_{\phi=0} < \mathcal{T}\,|_{\phi=\pi/2}
  \label{eq:requirementT}
\end{equation}
for $r>0$.  We note that $P_{l}^{2}(\cos\theta)$ has $(l-2)$ distinct zeros for $\cos\theta\in(-1,1)$~\cite{Henner2009}.  Since $\cos\theta \simeq (L^{2}-r^{2})/(L^{2}+r^{2})$ under the condition in Eq.~\eqref{eq:limitc}, $P_{l}^{2}(\cos\theta)$
exhibits $(l-2)$ distinct zeros for $r\in(0,\infty)$.  If modes with
$l>2$ dominated in Eq.~\eqref{eq:SolS}, then the physical constraint in Eq.~\eqref{eq:requirementT} would be violated as $P_{l>2}^{2}(\cos\theta)$ could change the sign at finite $r$.  
If the dominant component arises from $l=m=2$, the condition~\eqref{eq:requirementT} is guaranteed.

As long as the $l=2$ mode is sufficiently large, we can still consider higher
order perturbations originating from initial fluctuations.  Indeed, recent experimental measurements have observed spin polarization corresponding to the triangular flow, implying fluctuation-driven geometric deformations in non-central collisions~\cite{STAR:2023eck}.
We shall effectively incorporate the fluctuation effects by retaining the subdominant term with $m=3$ in Eq.~\eqref{eq:SolS}.  To ensure the monotonic thermal profile along the radial direction, we will focus on the mode with $l=m=3$ only.  Unlike the $m=2$ mode, symmetry properties cannot constrain the combination of $\cos(3\phi)$ and $\sin(3\phi)$.
We note that $c_{l3}\cos(3\phi)+d_{l3}\sin(3\phi)$ is proportional to $\cos[3(\phi+\phi_{0})]$ with a phase factor $\phi_{0}$.  This $\phi_{0}$ does not change the local spin polarization up to a phase factor, and the conclusions of this work do not depend on $\phi_0$.  Thus, we can safely set $\phi_{0}=0$, omitting all $\sin(3\phi)$ terms in Eq.~\eqref{eq:SolS}.

\begin{figure}
  \centering
  \includegraphics[width=0.98\columnwidth]{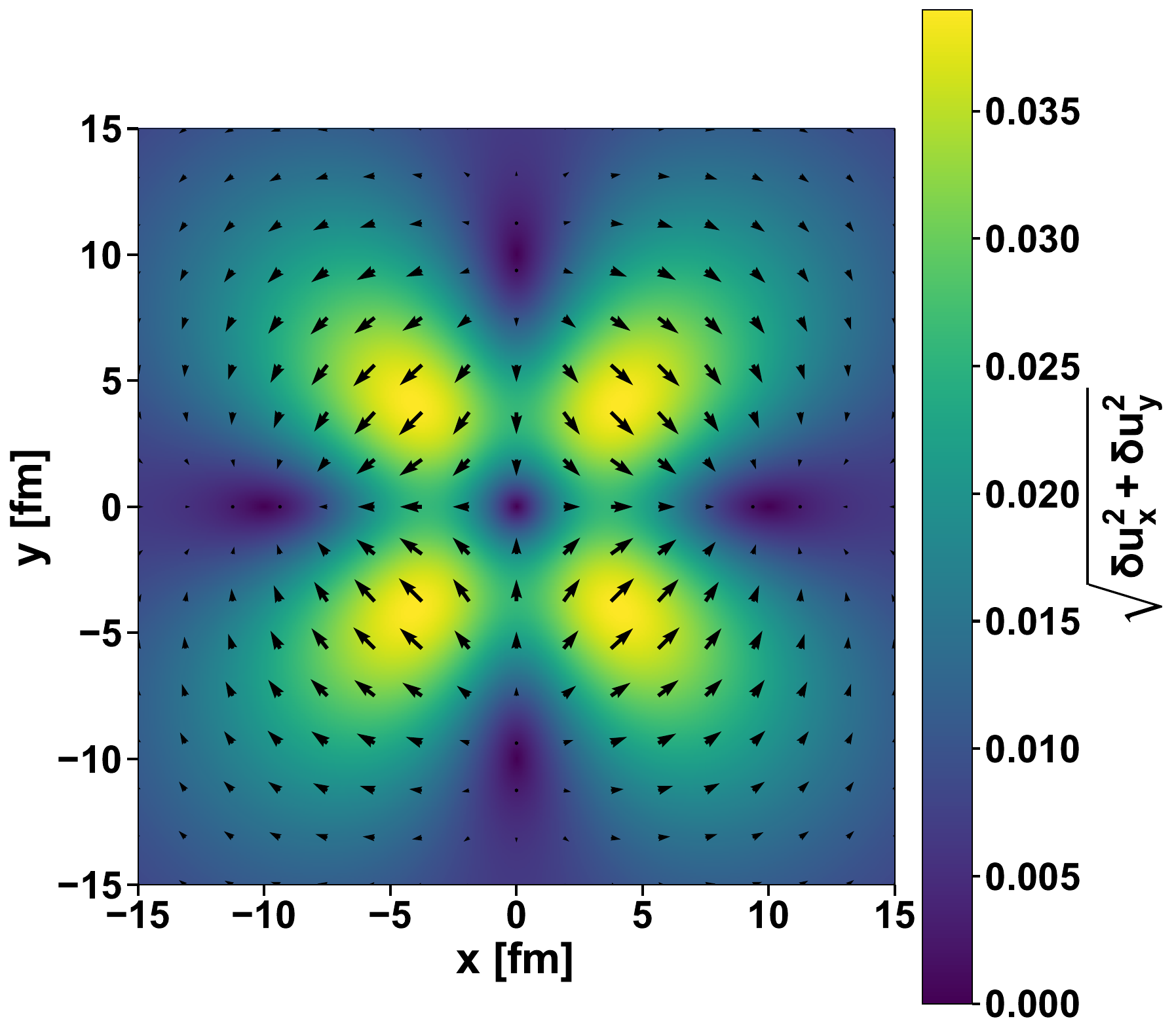}
  \caption{Transverse distribution of $(\delta u^{x}, {\delta u^{y}})$ in Eqs.~\eqref{eq:AnisotropicUx} and \eqref{eq:AnisotropicUy} with the mode coefficients, $\varepsilon_{2}=0.1$ and $\varepsilon_{3}=0$, and the physical setting, $L=\qty{10}{\femto\metre}$ and $\tau=\qty{1}{\femto\metre}$.  The arrows represent the direction of the velocity field, and the color indicates the magnitude of the velocity.  The velocity field exhibits a clear elliptic deformation.}
  \label{fig:du}
\end{figure}

Under the condition in Eq.~\eqref{eq:limitc}, the associated Legendre polynomial simplifies as follows:
\begin{equation}
  P_{l}^{l}(\cos\theta)\approx(-1)^{l}(2l-1)!!\frac{(2Lr)^{l}}{(L^{2}+r^{2})^{l}} \,.
  \label{eq:SolS2}
\end{equation}
Substituting Eqs.~\eqref{eq:approxForChiSigma}, \eqref{eq:SolS1}, \eqref{eq:SolS2} into Eqs.~\eqref{eq:sol} and transforming back to the Minkowski spacetime, we obtain the perturbed solution as~\cite{Hatta:2014jva}
\begin{subequations}
\begin{align}
  u_\tau & \simeq 1 \,,\qquad u_\eta = 0 \,,\\
  u_{r} & \simeq -\frac{\tau}{L}f(r)-3\varepsilon_{2}\tau f(r)f^{\prime}(r)\cos(2\phi) \notag\\
    & \qquad\quad-\frac{9}{2}\varepsilon_{3}\tau f^{2}(r)f^{\prime}(r)\cos(3\phi) \,,
    \label{eq:perturbedur}\\
  u_{\phi} & \simeq 3\varepsilon_{2}\tau f^{2}(r)\sin(2\phi)+\frac{9}{2}\varepsilon_{3}\tau f^{3}(r)\sin(3\phi)\;,
  \label{eq:perturbeduphi}
\end{align}
\label{eq:perturbedu}
\end{subequations}
with the corresponding temperature distribution given by
\begin{align}
 T & \simeq \frac{\hat{T}_{0}(2L)^{2/3}}{\tau^{1/3}(L^{2}+r^{2})^{2/3}}\left[1-\varepsilon_{2}f^{2}(r)\cos(2\phi)\right. \notag\\
    & \qquad\quad\left.-\varepsilon_{3}f^{3}(r)\cos(3\phi)\right] \,.
    \label{eq:perturbedT}
\end{align}
Here, $f^{\prime}(r)=df(r)/dr$ and 
\begin{equation}
  f(r)=\frac{2Lr}{L^{2}+r^{2}} \,.
\end{equation}
The new coefficients $\varepsilon_{2,3}$ are defined as $\varepsilon_{2}=-3c_{22}$
and $\varepsilon_{3}=15c_{33}$. To ensure the perturbative hierarchy between elliptic and triangular deformations, we shall impose the following condition:
\begin{equation}
0<\varepsilon_{3}\ll\varepsilon_{2}\ll1.
\end{equation}
Compared to the original Gubser flow, the perturbed solution introduces the $\phi$-dependence in both temperature and
velocity, describing the anisotropy in the transverse plane. For example, using $(t,x,y,z)$ coordinates, the anisotropic part related to $\varepsilon_{2}$ of the perturbed velocity is expressed as 
\begin{align}
  \delta u^{x} &= \frac{12\varepsilon_{2}L^{2}\tau x(L^{2}-x^{2}+3y^{2})}{(L^{2}+x^{2}+y^{2})^{3}} \,,
  \label{eq:AnisotropicUx}\\
  \delta u^{y} &= -\frac{12\varepsilon_{2}L^{2}\tau y(L^{2}+3x^{2}-y^{2})}{(L^{2}+x^{2}+y^{2})^{3}} \,,
  \label{eq:AnisotropicUy}
\end{align}
which are independent of the first term $-\tau f(r)/L$ in Eq.~\eqref{eq:perturbedur}.
The distribution of the anisotropic part of the perturbed velocity field is illustrated in Fig.~\ref{fig:du}.  The arrows represent the direction of the velocity field, and the color indicates the magnitude of the velocity.  We note that this perturbed velocity field exhibits a clear elliptic deformation, representing a finite elliptic-flow component.  The elliptic deformation by $\varepsilon_2$ is consistent with the geometry of non-central heavy-ion collisions, while the triangular deformation by $\varepsilon_3$ is induced by initial fluctuations.

We emphasize that the perturbed solution in Eqs.~\eqref{eq:perturbedu} and \eqref{eq:perturbedT} relies on the large size condition in Eq.~\eqref{eq:limitc}, which we assume holds even on the freeze-out hypersurface.  Denoting the freeze-out temperature as $\Tf$, we determine the freeze-out hypersurface by
\begin{equation}
  T(\tauf,r,\phi) = \Tf \,,
  \label{eq:FreezeoutHypersurface}
\end{equation}
Solving this equation yields the freeze-out proper time:
\begin{align}
  \tauf &\simeq \frac{4\hat{T}_{0}^{3}L^{2}}{\Tf^{3}(L^{2}+r^{2})^{2}}\left[1-3\varepsilon_{2}f^{2}(r)\cos(2\phi)\right.\notag \\
  & \qquad\quad\left.-3\varepsilon_{3}f^{3}(r)\cos(3\phi)\right]\,.
\end{align}
Consistency with the condition in Eq.~\eqref{eq:limitc} then requires
\begin{equation}
  \hat{T}^{3}_{0}/(\Tf L)^{3}\ll1\,,
  \label{eq:limitc1}
\end{equation}
which we assume throughout the subsequent analyses.

\section{Three Different Formulations}
\label{sec:Three-Different-Formulations}

The sign puzzle in local spin polarization suggests that the description based on thermal vorticity alone is incomplete, which has motivated the inclusion of shear-induced polarization.  Although the polarization induced by thermal vorticity is well-established, at least three distinct formulations for shear-induced polarization currently coexist in the literature.  We briefly review these formulations below.

The spin polarization of the $\Lambda$ hyperon in the Belinfante pseudo-gauge~\cite{belinfante1939spin,belinfante1940current,rosenfeld1940energy}, as a function of its
momentum, $p^\mu$, has two contributions, i.e., $S^\mu(p)=S_{\varpi}^\mu(p) + S_{\xi}^\mu(p)$, the former from the thermal vorticity tensor $\varpi_{\mu\nu}$ and the latter from the thermal shear tensor $\xi_{\mu\nu}$.  These contributions are calculated by the modified Cooper-Frye formula on the freeze-out hypersurface $\Sigma_\mathrm{f}$ as~\cite{Liu:2021uhn,Becattini:2021suc}
\begin{align}
  S_{\varpi}^{\mu}(p) &= \frac{-1}{8m\mathcal{N}(p)}\int_{\Sigma_\mathrm{f}}[\mathrm{d}\Sigma]\epsilon^{\mu\nu\rho\sigma}p_{\nu}\varpi_{\rho\sigma} \,,
  \label{eq:PolarizationFormulaVorticity}\\
  S_{\xi}^{\mu}(p) &= \frac{-1}{4m\mathcal{N}(p)}\int_{\Sigma_\mathrm{f}}[\mathrm{d}\Sigma]\epsilon^{\mu\nu\rho\sigma}\frac{p_{\nu}n_{\rho}}{p\cdot n}p^{\lambda}\xi_{\sigma\lambda}\,.
  \label{eq:PolarizationFormulaShear}
\end{align}
Before explaining the notation, we note that $p\cdot S_{\varpi,\xi}=0$ holds, and in the following, we work in the lab frame; we therefore focus on the spatial part, $\vec{S}$, only.

In the above expressions, we introduced a shorthand notation $[\mathrm{d}\Sigma]$ defined as $[\mathrm{d}\Sigma]=d\Sigma\cdot p \, n_{\rm F}(1-n_{\rm F})$, where $d\Sigma_{\mu}$ is the future-directed integration measure and $n_{\rm F}=(1+e^{\beta\cdot p})^{-1}$ is the Fermi-Dirac distribution function.
Since $e^{\beta\cdot p}\gg1$ in practical applications, we can approximate the weight factor as $n_{\rm F}(1-n_{\rm F})\simeq n_{\rm F}\simeq e^{-\beta\cdot p}$.

Here, the thermal vorticity tensor $\varpi_{\mu\nu}$ and thermal shear tensor $\xi_{\mu\nu}$ are given by
\begin{align}
  \varpi_{\mu\nu} & =-\frac{1}{2}(\partial_{\mu}\beta_{\nu}-\partial_{\nu}\beta_{\mu})\,,
  \label{eq:ThermalVorticity}\\
  \xi_{\mu\nu} & =\frac{1}{2}(\partial_{\mu}\beta_{\nu}+\partial_{\nu}\beta_{\mu})
  \label{eq:ThermalShearTensor}
\end{align}
with $\beta_{\mu}=u_{\mu}/T$. The normalization factor $\mathcal{N}(p)$ is calculated from
\begin{equation}
  \mathcal{N}(p)=\int_{\Sigma_\mathrm{f}}d\Sigma\cdot p \, n_{\rm F}\,.
  \label{eq:NormalizationNp}
\end{equation}

The choice of the unit vector $n^{\mu}$ in Eq.~\eqref{eq:PolarizationFormulaShear} may vary across different theoretical formulations.  In this work, we examine the following three distinct choices:
\begin{enumerate}
\item FLPSY formulation~\cite{Liu:2021uhn,Fu:2021pok}: $n^{\mu}$ is
identified with the fluid velocity $u^{\mu}$.
 
\item BBPIK formulation~\cite{Becattini:2021suc}: $n^{\mu}$ is set to be $n^{\mu}=\delta_{0}^{\mu}$ in $(t,x,y,z)$ coordinates.  In addition, as proposed in Ref.~\cite{Becattini:2021iol}, explicit $T$-gradients are removed by the replacements of $\varpi_{\mu\nu}\rightarrow\varpi_{\mu\nu}^{K}$
and $\xi_{\mu\nu}\rightarrow\xi_{\mu\nu}^{K}$, where 
\begin{align}
  \varpi_{\mu\nu}^{K} & =-\frac{1}{2}\beta(\partial_{\mu}u_{\nu}-\partial_{\nu}u_{\mu})\,,
  \label{eq:defomegak}\\
  \xi_{\mu\nu}^{K} & =\frac{1}{2}\beta(\partial_{\mu}u_{\nu}+\partial_{\nu}u_{\mu})\,.
  \label{eq:defsheark}
\end{align}

\item SBR formulation~\cite{Sheng:2025cjk}: $n^{\mu}$ is
the normal vector to $\Sigma_\mathrm{f}$. If $\Sigma_\mathrm{f}$ is isothermal, i.e., with constant temperature, $n^{\mu}=-\partial^{\mu}T/\sqrt{|\partial^{\nu}T\partial_{\nu}T|}$.  Then, $T$-gradient terms cancel out automatically in $S^{\mu}(p)$, though each $S_{\varpi,\xi}^{\mu}(p)$ receives $T$-gradient contributions.
\end{enumerate}
It should be noted that the SBR formulation is applicable to an arbitrary $\Sigma_\mathrm{f}$, and its complete form in Ref.~\cite{Sheng:2025cjk} is more general than Eqs.~\eqref{eq:PolarizationFormulaVorticity} and \eqref{eq:PolarizationFormulaShear}.  In the present work, we will calculate the polarization vector on an isothermal $\Sigma_\mathrm{f}$ determined by Eq.~\eqref{eq:FreezeoutHypersurface}.  In this way, we will omit unphysical contributions mentioned in Ref.~\cite{Sheng:2025cjk} and use formulas in Eqs.~\eqref{eq:PolarizationFormulaVorticity} and \eqref{eq:PolarizationFormulaShear} to evaluate the spin polarization in the SBR formulation.

For the mass $m$ appearing in Eqs.~\eqref{eq:PolarizationFormulaVorticity} and \eqref{eq:PolarizationFormulaShear}, the $\Lambda$ hyperon mass is the common choice, but as argued in Ref.~\cite{Fu:2021pok}, $m$ can be the $s$-quark mass if the $\Lambda$ spin polarization originates from the constituent $s$ quark.  In the subsequent  analyses, therefore, we will compare both scenarios: 
\begin{align}
  m = \begin{cases}
  1116\,\mathrm{MeV} & (\textrm{\ensuremath{\Lambda} hyperon}) \,,\\
  300\,\mathrm{MeV} & (\textrm{\ensuremath{s} quark}) \,.
  \end{cases}
  \label{eq:MassChoices}
\end{align}

\section{Spin Polarization with the Rotating Hubble Flow}
\label{sec:Spin-Polarization-Hubble}

We now calculate the spin polarization by using the analytical solutions presented in Sec.~\ref{sec:Analytical-Fluid-Solution}. Let us first consider the spin polarization in the rotating Hubble flow. For simplicity, we assume $|\vec{\omega}|\ll1$ and $|\vec{\omega} \times \vec{x}/t|<1$, 
and expand the relevant integrals in powers of $\vec{\omega}$.  Then, a nonzero polarization first appears at linear order in $\vec{\omega}$. In order to distinguish the results clearly, we will use superscripts FLPSY, BBPIK, and SBR to label three formulations shown in Sec.~\ref{sec:Three-Different-Formulations}.

In the FLPSY and the SBR formulations, the polarization induced by thermal vorticity is parallel to $\vec{\omega}$:
\begin{equation}
  \vec{S}_{\varpi}^{\mathrm{FLPSY}}(p) = \vec{S}_{\varpi}^{\mathrm{SBR}}(p) = \frac{E_{p}}{4mT_{0}}\vec{\omega} \,,
  \label{eq:HubbleSpinVor01}
\end{equation}
where $E_{p}=p^{0}$.  We note that $\vec{S}_{\xi}^{\mathrm{FLPSY}}(p) = \vec{S}_{\xi}^{\mathrm{SBR}}(p) = 0$ to all orders.  This is because the thermal shear tensor is proportional to the metric, i.e., $\xi_{\mu\nu}=T_{0}^{-1}g_{\mu\nu}$ in this case, making the integrand in Eq.~\eqref{eq:PolarizationFormulaShear} identically zero.  Therefore, the spin polarization receives only a vorticity-induced part, and no cancellation between $\vec{S}_{\varpi}$ and $\vec{S}_{\xi}$ can occur for the rotating Hubble flow.  This observation is consistent with the experimental data of global $\Lambda$ polarization, which is well described by the thermal vorticity alone~\cite{STAR:2017ckg}. 

In the BBPIK formulation, we obtain
\begin{align}
  \vec{S}_{\varpi}^{\mathrm{BBPIK}}(p) &= \frac{E_{p}}{4mT_{0}}\left[1+\frac{\Tf}{m}\frac{K_{3}(m/\Tf)}{K_{2}(m/\Tf)}\right]\vec{\omega}\,,
  \label{eq:HubbleSpinVor02}\\
  \vec{S}_{\xi}^{\mathrm{BBPIK}}(p) & =\frac{K_{4}(m/\Tf)}{8mT_{0}E_{p}K_{2}(m/\Tf)}[(\vec{\omega}\cdot\vec{p})\vec{p}-\vec{p}^{2}\vec{\omega}]\,,
  \label{eq:HubbleSpinVor03}
\end{align}
where $K_{\nu}(x)$ is the modified Bessel function of the second kind.  In a similar way to the FLPSY and the SBR formulations, $\vec{S}_{\varpi}^{\mathrm{BBPIK}}(p)$ remains parallel to $\vec{\omega}$.  There is a difference in the vorticity-induced part between the FLPSY/SBR results in Eq.~\eqref{eq:HubbleSpinVor01} and the BBPIK results in Eq.~\eqref{eq:HubbleSpinVor02}.  This difference is a function of $m/\Tf$, and it decreases with increasing $m/\Tf$.  The shear-induced polarization $\vec{S}_{\xi}^{\mathrm{BBPIK}}$ is always perpendicular to $\vec{p}$.  
For global polarization, we average Eqs.~\eqref{eq:HubbleSpinVor02} and \eqref{eq:HubbleSpinVor03} over the momentum $\vec{p}$. Upon averaging, the direction of $\vec{S}_{\varpi}^{\mathrm{BBPIK}}$ remains unchanged, while $\vec{S}_{\xi}^{\mathrm{BBPIK}}$ becomes strictly anti-parallel to $\vec{S}_{\varpi}^{\mathrm{BBPIK}}$. The two contributions therefore partially cancel, yet the net polarization remains aligned with $\vec{\omega}$.
We emphasize that, for global polarization, the shear-induced polarization contributions in the BBPIK formula remains nonzero.


\section{Spin Polarization with the Perturbed Gubser Flow \label{sec:Spin-Polarization-from}}

In this section, we consider the spin polarization using the perturbed Gubser flow given in Eqs.~\eqref{eq:perturbedu} and \eqref{eq:perturbedT}.

\subsection{Analytical Expressions of the Spin Polarization}

We parametrize the momentum $p^{\mu}=(E_{p},p^{x},p^{y},p^{z})$ as
follows:
\begin{align}
  \begin{split}
  & E_{p} = m_{\mathrm{T}}\cosh\eta_{p} \,, \qquad
    p^{z} = m_{\mathrm{T}}\sinh\eta_{p} \,,\\
  & p^{x} = p_{\mathrm{T}}\cos\phi_{p} \,,\qquad
    p^{y} = p_{\mathrm{T}}\sin\phi_{p} \,,
  \end{split}
\end{align}
where, conversely, we have $\eta_{p}=\tanh^{-1}(p^{z}/E_{p})$, $\phi_{p}=\tan^{-1}(p^{y}/p^{x})$, $p_{\mathrm{T}}=(p_{x}^{2}+p_{y}^{2})^{1/2}$, and $m_{\mathrm{T}}=(p_{\mathrm{T}}^{2}+m^{2})^{1/2}$.
With this parametrization, the integration measure $d\Sigma\cdot p$
on the freeze-out hypersurface $\Sigma_\mathrm{f}$ can be expressed as
\begin{multline}
  d\Sigma \cdot p = dr\, d\phi\, d\eta\, \tauf [ r m_{\mathrm{T}} \cosh(\eta_{p}-\eta) \\
  - r p_{\mathrm{T}} \cos(\phi_{p}-\phi) \partial_{r}\tauf 
  - p_{\mathrm{T}} \sin(\phi_{p}-\phi) \partial_{\phi}\tauf ]
  \label{eq:3}
\end{multline}
with integration domains $r\in[0,+\infty)$, $\phi\in[0,2\pi)$, and
$\eta\in(-\infty,+\infty)$.

After evaluating the integrals in Eqs.~\eqref{eq:PolarizationFormulaVorticity} and \eqref{eq:PolarizationFormulaShear}, we will average the results over a symmetric interval of $\eta_p$ centered at the origin, following the experimental procedure.  Then, $S^{x} = S^{y} = 0$, because the integrands are odd under $(\eta, \eta_p) \to (-\eta, -\eta_p)$, reflecting the underlying symmetry of the collision system.  Thus, only the longitudinal component, $S^{z}(p)$, survives in the averaging process, which significantly simplifies our subsequent analyses.

To derive analytical expressions of the longitudinal spin polarization, we expand the integrals for small $\varepsilon_{2,3}$ and large $L$.  For the normalization factor in Eq.~\eqref{eq:NormalizationNp}, the leading order term is given by
\begin{equation}
  \mathcal{N}(p) = \frac{8\pi\hat{T}_{0}^{3}\vartheta}{\Tf^{2}}K_{1}(\vartheta) \,,
  \qquad
  \vartheta=\frac{\sqrt{m^{2}+p_{\mathrm{T}}^{2}}}{\Tf} \,.
  \label{eq:thetaDef}
\end{equation}
For the longitudinal spin polarization $S^{z}$, we first take an expansion in Eqs.~\eqref{eq:PolarizationFormulaVorticity} and \eqref{eq:PolarizationFormulaShear} in powers of $\varepsilon_{2,3}$ up to the linear order:
\begin{align}
  S_{\varpi}^{z}(p) &= \sum_{i=2}^{3}\varepsilon_{i}\mathcal{S}_{(i),\varpi}(p)\sin(i\phi_{p}) \,,
  \label{eq:expansionSomega0}\\
  S_{\xi}^{z}(p) &= \sum_{i=2}^{3}\varepsilon_{i}\mathcal{S}_{(i),\xi}(p)\sin(i\phi_{p}) \,.
  \label{eq:expansionSxi0}
\end{align}
Then, we expand the Fourier coefficients, $\mathcal{S}_{(2)}(p)$ and $\mathcal{S}_{(3)}(p)$, for large $L$ such that the condition in
Eq.~\eqref{eq:limitc1} is satisfied.  We keep the leading order terms correspondingly, by changing the integration variable as $r\to s=r/L$ in Eqs.~\eqref{eq:PolarizationFormulaVorticity} and \eqref{eq:PolarizationFormulaShear}, and expanding the integrands in powers of $1/(\Tf L)$.  We then perform the integrations over $\phi\in[0,2\pi)$, $s\in[0,+\infty)$,
and $\eta\in(-\infty,+\infty)$. 

The analytical results are summarized as follows.  For the polarization induced by thermal vorticity, we find the relation:
\begin{equation}
  \mathcal{S}_{(i),\varpi}^{\mathrm{FLPSY}}(p) = 2\mathcal{S}_{(i),\varpi}^{\mathrm{BBPIK}}(p) = \mathcal{S}_{(i),\varpi}^{\mathrm{SBR}}(p)
  \label{eq:SvorticityRelation}
\end{equation}
for $i=2,3$, and
\begin{align}
  \mathcal{S}_{(2),\varpi}^{\mathrm{SBR}}(p) &= -\mathcal{C}(p)[\vartheta K_{0}(\vartheta) + 3K_{1}(\vartheta)] \,,\\
  \mathcal{S}_{(3),\varpi}^{\mathrm{SBR}}(p) &= -\mathcal{D}(p)[\vartheta K_{0}(\vartheta) + 5K_{1}(\vartheta)] \,,
\end{align}
where the coefficients, $\mathcal{C}(p)$ and $\mathcal{D}(p)$, read:
\begin{align}
  \mathcal{C}(p) &= \frac{96\pi\hat{T}_{0}^{6}p_{\mathrm{T}}^{2}\cosh\eta_{p}}{35m \Tf^{7}L^{4}\mathcal{N}(p)} \,,
  \label{eq:defCp}\\
  \mathcal{D}(p) &= \frac{144\pi\hat{T}_{0}^{9}p_{\mathrm{T}}^{3}\cosh\eta_{p}}{55m \Tf^{11}L^{7}\mathcal{N}(p)} \,.
  \label{eq:defDp}
\end{align}
For large $L$, we find $\mathcal{C}(p)\gg\mathcal{D}(p)$, which is consistent
with the assumption that the elliptic deformation is dominant.

We note that in Eq.~\eqref{eq:SvorticityRelation} $\mathcal{S}_{(i),\varpi}^{\mathrm{BBPIK}}(p)$ is half of $\mathcal{S}_{(i),\varpi}^{\mathrm{FLPSY}}$ and $\mathcal{S}_{(i),\varpi}^{\mathrm{SBR}}$.  The perturbed solution in Eqs.~\eqref{eq:perturbedu} and \eqref{eq:perturbedT} satisfies 
\begin{align}
  \varpi_{\mu\nu} &= -\frac{\beta}{2}(\partial_{\rho}u_{\sigma}-\partial_{\sigma}u_{\rho}) - \frac{1}{2}(u_{\sigma}\partial_{\rho}\beta-u_{\rho}\partial_{\sigma}\beta)\notag \\
  &= -\beta(\partial_{\rho}u_{\sigma}-\partial_{\sigma}u_{\rho}) + \mathcal{O}(\varepsilon_{2,3}^{2}) \,,
\end{align}
while $\varpi_{\mu\nu}$ is replaced with $-\frac{1}{2}\beta(\partial_{\rho}u_{\sigma}-\partial_{\sigma}u_{\rho})$ in the BBPIK formulation as indicated in Eq.~\eqref{eq:defomegak}.  This explains the factor of $1/2$ in Eq.~\eqref{eq:SvorticityRelation}~\cite{Cao}.

\begin{figure*}[t]
\begin{centering}
\includegraphics[scale=0.5]{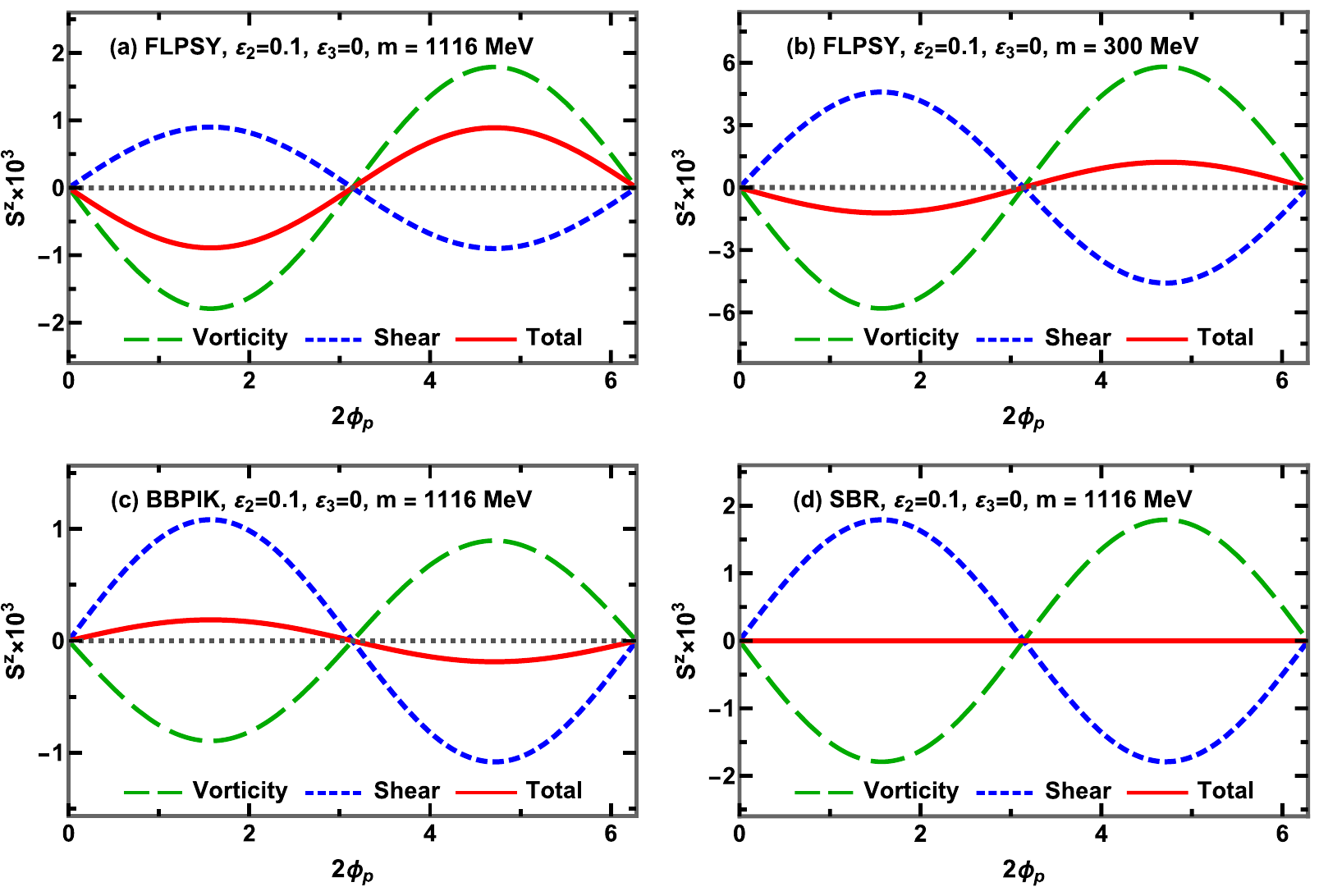}
\par\end{centering}
\caption{Longitudinal spin polarization with $\varepsilon_{2}=0.1$ and $\varepsilon_{3}=0$
as a function of the azimuthal angle $\phi_{p}$.  The green dashed
line represents $S_{\varpi}^{z}$, the blue dashed line represents
$S_{\xi}^{z}$, and the red solid line represents the total polarization.
The panels display results for different formulations: (a) and (b) for FLPSY, (c) for BBPIK, and (d) for SBR\@.
The mass $m$ is set to $\qty{300}{\mathrm{MeV}}$ in panel (b) and $\qty{1116}{\mathrm{MeV}}$ in others.  The remaining parameters are fixed as $\hat{T}_{0}=3$,
$L=\qty{10}{\femto\metre}$, and $\Tf=\qty{150}{\mathrm{MeV}}$.  The results
are averaged over the rapidity range $\eta_{p}\in[-1,1]$ and the
transverse momentum range $p_{\mathrm{T}}\in[0.5,3]\,\mathrm{GeV}$.
\label{fig:Longitudinal-spin-polarization-2}}
\end{figure*}

For the polarization induced by thermal shear tensor, we obtain:
\begin{align}
  \mathcal{S}_{(2),\xi}^{\mathrm{FLPSY}}(p) &= 3\mathcal{C}(p)\left[ 3K_{1}(\vartheta) - \int_{\vartheta}^{\infty}dt\,K_{0}(t)\right] \,,\\
  \mathcal{S}_{(3),\xi}^{\mathrm{FLPSY}}(p) &= 3\mathcal{D}(p)\left[ 10\vartheta^{-1}K_{0}(\vartheta) + 3K_{1}(\vartheta)\right]
\end{align}
for the FLPSY formulation, and
\begin{align}
  \mathcal{S}_{(2),\xi}^{\mathrm{BBPIK}}(p) &= \frac{\mathcal{C}(p)}{2\cosh^{2}\eta_{p}}[\vartheta K_{0}(\vartheta) + 11K_{1}(\vartheta)] \,,\\
  \mathcal{S}_{(3),\xi}^{\mathrm{BBPIK}}(p) &= \frac{\mathcal{D}(p)}{2\cosh^{2}\eta_{p}}[17K_{1}(\vartheta) + (\vartheta+24\vartheta^{-1})K_{0}(\vartheta)]
\end{align}
for the BBPIK formulation, and
\begin{equation}
  \mathcal{S}_{(2,3),\xi}^{\mathrm{SBR}}(p) = -\mathcal{S}_{(2,3),\varpi}^{\mathrm{SBR}}(p)
\end{equation}
for the SBR formulation.

For large $L$, in all the cases, $\mathcal{S}_{(i),\xi}(p)$ turns out to be of the same order as $\mathcal{S}_{(i),\varpi}(p)$.  In particular for the SBR formulation, they cancel exactly and the total polarization vanishes.


\subsection{Dependence on the Azimuthal Angle and the Transverse Momenta}

\begin{figure*}
\begin{centering}
\includegraphics[scale=0.5]{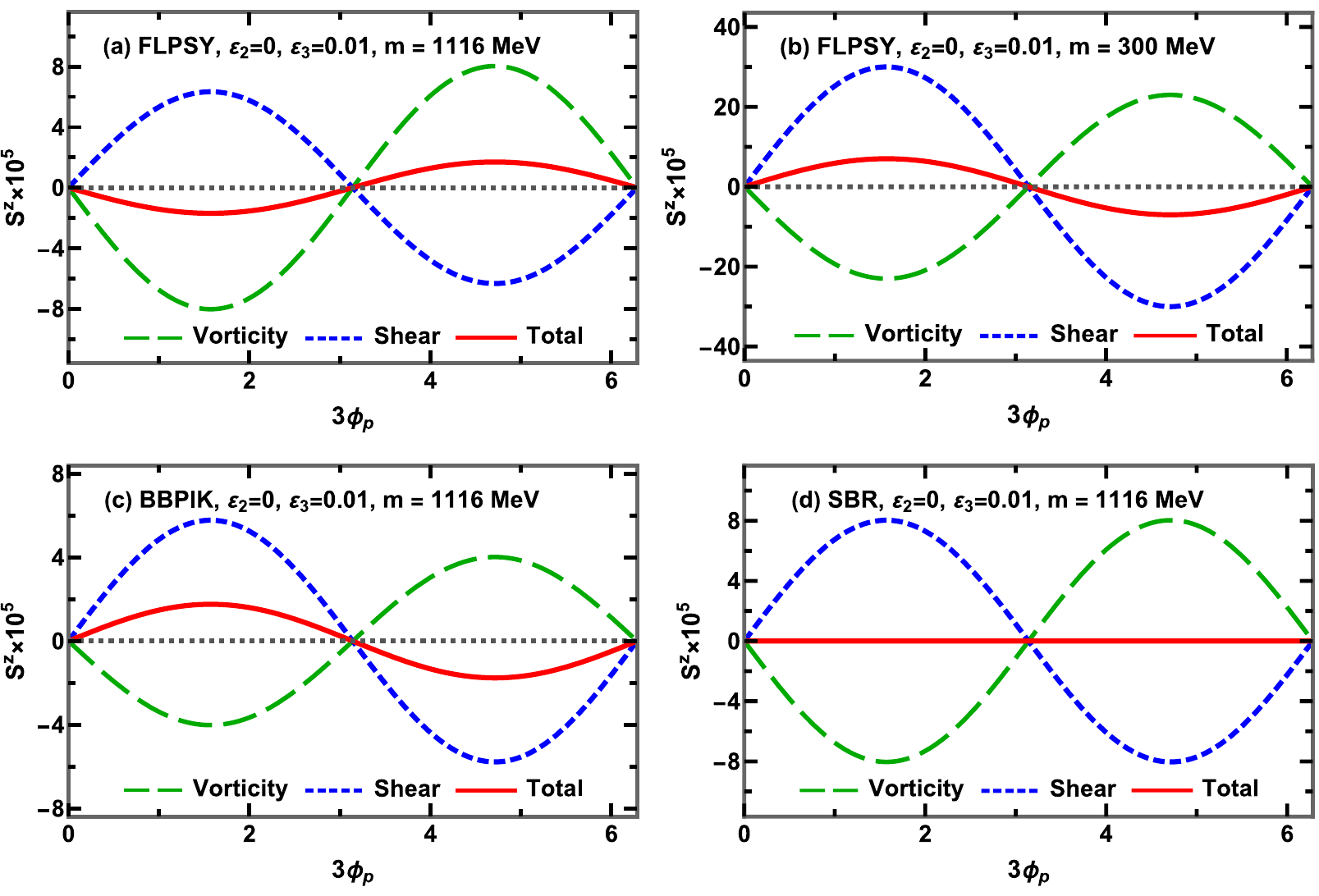}
\par\end{centering}
\caption{Longitudinal spin polarization with $\varepsilon_{2}=0$ and $\varepsilon_{3}=0.01$ as a function of the azimuthal angle $\phi_{p}$.  The green dashed
line represents $S_{\varpi}^{z}$, the blue dashed line represents
$S_{\xi}^{z}$, and the red solid line represents the total polarization.
The panels display results for different formulations: (a) and (b) for FLPSY, (c) for BBPIK, and (d) for SBR\@.
The mass $m$ is set to $\qty{300}{\mathrm{MeV}}$ in panel (b) and $\qty{1116}{\mathrm{MeV}}$ in others.  The remaining parameters are fixed as $\hat{T}_{0}=3$,
$L=\qty{10}{\femto\metre}$, and $\Tf=\qty{150}{\mathrm{MeV}}$.  The results
are averaged over the rapidity range $\eta_{p}\in[-1,1]$ and the
transverse momentum range $p_{\mathrm{T}}\in[0.5,3]\,\mathrm{GeV}$.
\label{fig:Longitudinal-spin-polarization-3}}
\end{figure*}

Let us first examine the dependence on the azimuthal angle $\phi_{p}$.  For illustration, $S_{\varpi}^{z}$, $S_{\xi}^{z}$, and their sum are plotted as functions of $\phi_{p}$ in Figs.~\ref{fig:Longitudinal-spin-polarization-2} and \ref{fig:Longitudinal-spin-polarization-3}.  We set $\varepsilon_{2}=0.1$ and $\varepsilon_{3}=0$ to isolate the $\sin(2\phi_{p})$ structure in Fig.~\ref{fig:Longitudinal-spin-polarization-2}, and $\varepsilon_{2}=0$ and $\varepsilon_{3}=0.01$ to extract the $\sin(3\phi_{p})$ structure in Fig.~\ref{fig:Longitudinal-spin-polarization-3}.  Other parameters are chosen as $\hat{T}_{0}=3$, $L=\qty{10}{\femto\metre}$, and $\Tf=\qty{150}{\mathrm{MeV}}$.  The results
are averaged over the rapidity range, $\eta_{p}\in[-1,1]$ and the transverse
momentum, $p_{\mathrm{T}}\in[0.5,3]\,\mathrm{GeV}$. 

From Figs.~\ref{fig:Longitudinal-spin-polarization-2} and \ref{fig:Longitudinal-spin-polarization-3}, we see that $S_{\xi}^{z}$ has the desired sign consistent with the experimental data~\citep{STAR:2019erd,STAR:2023eck}, while $S_{\varpi}^{z}$ always has the opposite sign.  This can be understood from our formulas with $\mathcal{C}(p), \mathcal{D}(p) > 0$ in Eqs.~\eqref{eq:defCp} and \eqref{eq:defDp}.  To reproduce the desired sign of total longitudinal spin polarization, the magnitude of the thermal-shear contribution should exceed that of thermal vorticity.  For panels (a) and (b) of Fig.~\ref{fig:Longitudinal-spin-polarization-2} in the FLPSY formulation, the total spin polarization shown by the solid line has the undesired sign regardless of the mass choice. 
Interestingly, we find that the total longitudinal spin polarization in the FLPSY formalism becomes positive in panel (b) of Fig.~\ref{fig:Longitudinal-spin-polarization-3}.
In the SBR formulation in panel (d) of Figs.~\ref{fig:Longitudinal-spin-polarization-2} and~\ref{fig:Longitudinal-spin-polarization-3}, the total polarization is identically zero as a result of exact cancellation at leading order. The BBPIK formulation in panel (c) of Figs.~\ref{fig:Longitudinal-spin-polarization-2} and~\ref{fig:Longitudinal-spin-polarization-3} produces the desired sign, arising after a significant cancellation between $S_{\varpi}^{z}$ and $S_{\xi}^{z}$.  For the BBPIK and SBR formulations, here, we present only the results for $m=\qty{1116}{\mathrm{MeV}}$; we have numerically checked that the sign in the BBPIK formulation remains unchanged for $m=\qty{300}{\mathrm{MeV}}$.

As another interesting observation, we find that the longitudinal spin polarization vanishes when $\epsilon_2=\epsilon_3=0$. This indicates that finite initial eccentricities generate anisotropic flow, which in turn leads to longitudinal spin polarization. Therefore, initial-state fluctuations can influence longitudinal spin polarization.

\begin{figure}
\begin{centering}
\includegraphics[width=0.9\columnwidth]{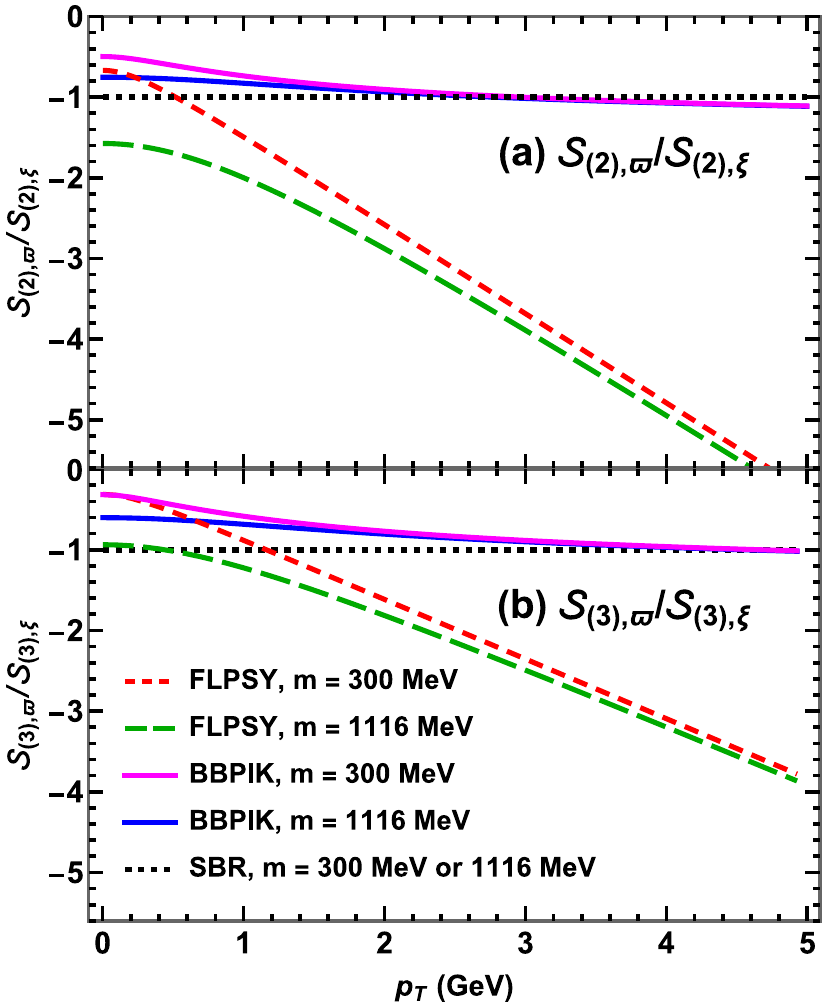}
\par\end{centering}
\caption{$\mathcal{S}_{(i),\varpi}/\mathcal{S}_{(i),\xi}$ as a function
of the transverse momentum $p_{\mathrm{T}}$.  Panel (a) shows $\mathcal{S}_{(2),\varpi}/\mathcal{S}_{(2),\xi}$ $(i=2)$, while
panel (b) shows $\mathcal{S}_{(3),\varpi}/\mathcal{S}_{(3),\xi}$ ($i=3$).
The freeze-out temperature is taken as $\Tf=\qty{150}{\mathrm{MeV}}$.
$\mathcal{S}_{(i),\varpi}$ and $\mathcal{S}_{(i),\xi}$ are separately averaged over $\eta_{p}\in[-1,1]$.}
\label{fig:TheRatio}
\end{figure}

Next, we shall discuss the dependence on the transverse momentum $p_{\mathrm{T}}$.  Since the relative magnitudes of $S_{\varpi}^{z}$ and $S_{\xi}^{z}$ determine the sign of their sum, we plot the ratio, $\mathcal{S}_{(i),\varpi}/\mathcal{S}_{(i),\xi}$, averaged over $\eta_{p}\in[-1,1]$ as a function of $p_{\mathrm{T}}$ in Fig.~\ref{fig:TheRatio}.  Interestingly, this ratio is independent of $\hat{T}_{0}$ and $L$, implying that both $S_{\varpi}^{z}$ and $S_{\xi}^{z}$ identically respond to the initial conditions.  Obtaining the experimentally desired sign for the total longitudinal spin polarization requires $\mathcal{S}_{(i),\varpi}/\mathcal{S}_{(i),\xi}>-1$.

Comparing the results for $m=\qty{1116}{\mathrm{MeV}}$ and $m=\qty{300}{\mathrm{MeV}}$, we find that a smaller mass yields a larger ratio, $\mathcal{S}_{(i),\varpi}/\mathcal{S}_{(i),\xi}$, which can be understood from the $(p_\nu n_\rho)/(p \cdot n)$ factor in $S^\mu_\xi$~\cite{Cao}.  In particular, for the FLPSY formulation, Fig.~\ref{fig:TheRatio} shows $\mathcal{S}_{(i),\varpi}^{\mathrm{FLPSY}}/\mathcal{S}_{(i),\xi}^{\mathrm{FLPSY}} > -1$ for small $p_{\mathrm{T}}$ in the case of $m=\qty{300}{\mathrm{MeV}}$.  Although the ratio lies above $-1$ in the small-$p_{\mathrm{T}}$ region giving the correct sign there, the ratio decays rapidly as $p_{\mathrm{T}}$ increases and becomes much less than $-1$ at large $p_{\mathrm{T}}$.
 This behavior is consistent with the results in Figs.~\ref{fig:Longitudinal-spin-polarization-2} and \ref{fig:Longitudinal-spin-polarization-3}, where the spin polarization averaged over $p_{\mathrm{T}}\in[0.5,3]\,\mathrm{GeV}$ has an undesired sign.

The BBPIK formulation works well when $p_{\mathrm{T}}<3\,\mathrm{GeV}$, as shown in Fig.~\ref{fig:TheRatio}.  As $p_{\mathrm{T}}$ increases, the ratio decreases slowly but eventually crosses $-1$.  At large $p_{\mathrm{T}}$, we find $\mathcal{S}_{(i),\varpi}^{\mathrm{BBPIK}}/\mathcal{S}_{(i),\xi}^{\mathrm{BBPIK}}<-1$, which suggests that the BBPIK formulation also predicts the opposite sign for the longitudinal spin polarization at large transverse momenta.

\subsection{Acceleration-driven Near Cancellation}

From Figs.~\ref{fig:Longitudinal-spin-polarization-2} -~\ref{fig:TheRatio}, we find a significant cancellation between $S_{\varpi}^{\mu}$ and $S_{\xi}^{\mu}$ in the BBPIK formulation.  This cancellation can partially be understood by comparing the integrands of $S_{\varpi}^{z}(p)$ in Eq.~\eqref{eq:PolarizationFormulaVorticity} and $S_{\xi}^{z}(p)$ in Eq.~\eqref{eq:PolarizationFormulaShear}.  Recalling that the replacements, $\varpi_{\mu\nu}\rightarrow\varpi_{\mu\nu}^{K}$ and $\xi_{\mu\nu}\rightarrow\xi_{\mu\nu}^{K}$, are adopted in the BBPIK formulation, we then find 
\begin{align}
  \frac{1}{2}\epsilon^{z\nu\rho\sigma}p_{\nu}\varpi_{\rho\sigma}^{K}
    &= p_{x}\varpi_{ty}^{K} - p_{y}\varpi_{tx}^{K} + \cdots \,,
    \label{eq:cancel01}\\
  \epsilon^{z\nu\rho\sigma}\frac{p_{\nu}n_{\rho}}{p\cdot n}p^{\lambda}\xi_{\sigma\lambda}^{K}
    &= p_{x}\xi_{ty}^{K} - p_{y}\xi_{tx}^{K} + \cdots \,.
    \label{eq:cancel02}
\end{align}
Here, the ellipses represent the contributions from other components
of $\varpi_{\mu\nu}^{K}$ and $\xi_{\mu\nu}^{K}$.  We include a factor of $1/2$ in Eq.~\eqref{eq:cancel01} to account for the factor-two difference between the denominators in Eqs.~\eqref{eq:PolarizationFormulaVorticity} and \eqref{eq:PolarizationFormulaShear}.
According to the definitions of $\varpi_{\mu\nu}^{K}$ and $\xi_{\mu\nu}^{K}$,
the acceleration terms $\partial_{t}u_{x}$ and $\partial_{t}u_{y}$
cancel exactly in the sum of two contributions in Eqs.~\eqref{eq:cancel01} and \eqref{eq:cancel02}~\cite{Cao}.  Therefore, the acceleration effect does not cause the longitudinal spin polarization in the BBPIK formulation.  We emphasize that this argument does not rely on any specific velocity profile.

In the perturbed Gubser flow given by Eq.~\eqref{eq:perturbedu}, the terms represented by the ellipses in Eq.~\eqref{eq:cancel01} vanish, and $\varpi_{tx}^{K}=-\xi_{tx}^{K}$ and $\varpi_{ty}^{K}=-\xi_{ty}^{K}$ immediately follow.  This implies that $S_{\varpi}^{z}(p)$ is completely eliminated by acceleration part of $S_{\xi}^{z}(p)$ in the BBPIK formulation.  In this specific case, the total longitudinal spin polarization depends only on the non-acceleration components
of $\xi_{\mu\nu}$ but not $\varpi_{\mu\nu}$. 

The cancellation of the acceleration terms also occurs in the SBR formulation.  In the QGP generated in relativistic heavy-ion collisions, the longitudinal expansion dominates over the transverse expansion.  If $\eta$-dependence is negligible, the anisotropic expansion leads to the following condition for the temperature gradients:
\begin{equation}
  |\partial_{x}T|\,,\; |\partial_{y}T| ~\ll~ |\partial_{z}T| \,.
  \label{eq:ConditionTgradient}
\end{equation}
The unit vector, $n^{\mu}=-\partial^{\mu}T/\sqrt{|\partial^{\nu}T\partial_{\nu}T|}$, in the SBR formulation can thus be approximated as
\begin{equation}
  (n^{t},n^{x},n^{y},n^{z}) \simeq \frac{(-\partial_{t}T,0,0,\partial_{z}T)}{\sqrt{(\partial_{t}T)^{2}-(\partial_{z}T)^{2}}} \,.
\end{equation}
Since $\partial_{t}T=-\partial_{z}T\coth\eta$ due to $\partial_\eta T=0$, we can write the above expression as
\begin{equation}
  n^{\mu} = \delta_{0}^{\mu} + \delta n^{\mu} \,,
\end{equation}
where $\delta n^{t}=\cosh\eta-1$, $\delta n^{z}=\sinh\eta$, $\delta n^{x} \simeq 0$, $\delta n^{y} \simeq 0$ under the condition in Eq.~\eqref{eq:ConditionTgradient}.
In the mid rapidity region $\eta_{p}\approx 0$, the distribution function,
$n_{\rm F}$, provides an exponential suppression factor $e^{-\vartheta\cosh\eta}$.
Hence, the dominant contribution to the integral in Eq.~\eqref{eq:PolarizationFormulaShear} comes from the region where $\eta\approx 0$, and thus $\delta n^{t}\approx\delta n^{z}\approx 0$.
The vector $n^{\mu}$ can now be approximated as $\delta_{0}^{\mu}$.
Thanks to the exponential suppression, this approximation becomes increasingly accurate in the SBR formulation as $\vartheta$ increases, i.e., $p_{\mathrm{T}}$ increases.  Recall that, in the SBR formulation, the replacement of $\varpi_{\mu\nu} \to \varpi_{\mu\nu}^{K}$ and $\xi_{\mu\nu} \to \xi_{\mu\nu}^{K}$ does not change the total polarization vector $S^{z}(p)$~\citep{Sheng:2025cjk}. Therefore, Eqs.~\eqref{eq:cancel01} and \eqref{eq:cancel02} are approximately applicable to the SBR formulation, so that the acceleration terms
in $S_{\varpi}^{z}$ and $S_{\xi}^{z}$ cancel each other, as in the
BBPIK formulation.

Finally, we make a remark about the total longitudinal spin polarization in the SBR formulation, which is exactly vanishing as shown in Fig.~\ref{fig:TheRatio}, which still lacks a theoretical explanation, and is probably an accidental cancellation.
As discussed above, $S_{\varpi}^{z}$ is completely eliminated by the acceleration
part of $S_{\xi}^{z}$ in the perturbed Gubser flow.  The exact cancellation requires that the non-acceleration part of $S_{\xi}^{z}$ should also vanish.
We have numerically verified that the integrand of the complicated non-acceleration
part of $S_{\xi}^{z}$ is nonzero, while the integration over the full hypersurface
makes the final result vanish.  We have also examined the effect of introducing
finite cutoffs for $\eta$ and $r$ in the integration.  Although this renders total longitudinal spin polarization finite, its magnitude remains much smaller than $S_{\varpi}^{z}$ and $S_{\xi}^{z}$.

\section{Conclusion \label{sec:Conclusion}}

In this work, we have analytically studied the longitudinal spin polarization of $\Lambda$ hyperons in relativistic heavy-ion collisions using analytical hydrodynamic solutions.  Based on the rotating Hubble flow and the perturbed Gubser flow including elliptic and triangular fluctuations, we have derived analytical expressions for the spin polarization vector in three distinct ways: the FLPSY, BBPIK, and SBR formulations.  Our analytical approach provides a useful insight to investigate the thermal-vorticity and thermal-shear contributions to the longitudinal spin polarization. 

In the rotating Hubble flow, which could qualitatively be regarded as an approximation of the global polarization setup, the polarization induced by thermal vorticity is found to be parallel to the angular velocity vector $\vec{\omega}$ in all formulations under consideration.  The contribution from the thermal shear tensor vanishes identically in the FLPSY and SBR formulations, while in the BBPIK formulation it becomes nonzero and contains a component antiparallel to $\vec{\omega}$, leading to the tendency of partial cancellation. For global polarization, the shear-induced polarization in the BBPIK formulation becomes strictly antiparallel to $\vec{\omega}$ after an average over spatial momentum, yet the net polarization remains aliged with $\vec{\omega}$. 

In perturbed Gubser flow, in contrast, our results have revealed that, across all evaluated formulations, the thermal vorticity alone always yields a sign opposite to the experimental observations.  This motivates the inclusion of thermal shear tensor, which is expected to recover the desired overall sign.  However, the final result is highly sensitive to the chosen formulation.  In the FLPSY formulation, we have found that the experimentally favored sign of the longitudinal spin polarization is recovered only when the $s$ quark mass ($m=\qty{300}{\mathrm{MeV}}$) is used, and only within the small transverse momentum regime.  In the BBPIK formulation, the predicted total longitudinal polarization has a correct sign.  In the SBR formulation, it vanishes at leading order after integration over the full isothermal freeze-out hypersurface.

We have found that the acceleration terms cancel in both the BBPIK and SBR formulations.  We have demonstrated that the acceleration terms arising from the thermal vorticity are canceled by the acceleration terms in the thermal shear tensor.  Consequently, the longitudinal spin polarization is determined primarily by the non-acceleration effects.  This cancellation partly explains why the net polarization is significantly smaller than the individual thermal-vorticity and thermal-shear contributions.  We emphasize that such cancellation does not rely on velocity profiles.

In the present work, our aim was to deepen our understanding about the roles of the acceleration and non-acceleration terms in the thermal-vorticity and thermal-shear contributions at the analytical level.  Thus, we did not try to adjust the parameters, $T_0$, $L$, $\varepsilon_{2,3}$, to fit the experimental data.  However, the perturbed Gubser flow could be a reasonable approximation at sufficiently high collision energies as measured at LHC~\cite{ALICE:2021pzu}.  Then, in the future, it is worth extending our analytical approach for phenomenological study. For this purpose, it would be crucial to relax the large-size condition in Eq.~\eqref{eq:limitc}.

Future studies could also generalize our consideration to more realistic solutions of relativistic hydrodynamics including viscous terms to study the spin polarization, uncovering or confirming approximate cancellation mechanisms.

\begin{acknowledgments}
The authors are grateful to Zheng~Cao for useful discussions particularly in the early stage of this work.
The authors thank
Xu-Guang~Huang
and
Yi~Yin
for helpful discussions.
KF is supported in part by Japan Society for the Promotion of Science (JSPS) KAKENHI Grant Nos.\ 22H05118, 25K24464, 26K00698.
SP is supported in part by the National Key Research and Development Program of China under Contract No.\ 2022YFA1605500,
by the Chinese Academy of Sciences (CAS) under Grant No.\ YSBR-088,
and by the National Natural Science Foundation of China (NSFC) under Grant No.\ 12135011.
DLW is supported in part by the NSFC under Grant No.\ 125B2110.
\end{acknowledgments}

\bibliography{refs}

@unpublished{Ren:2026fqj,
    author = "Ren, Xiang and Hu, Jin-Yu and Xu, Hao-jie and Pu, Shi",
    title = "{Nonlinear response of flow harmonics in Gubser flow with participant-reaction planes mismatch}",
    eprint = "2604.11058",
    archivePrefix = "arXiv",
    primaryClass = "nucl-th",
    month = "4",
    year = "2026"
}

@article{belinfante1939spin,
  author  = {Belinfante, F. J.},
  journal = {Physica},
  volume  = {6},
  pages   = {887},
  year    = {1939}
}

@article{belinfante1940current,
  author  = {Belinfante, F. J.},
  journal = {Physica},
  volume  = {7},
  pages   = {449},
  year    = {1940}
}

@article{rosenfeld1940energy,
  author  = {Rosenfeld, L.},
  journal = {M{\'e}m. Acad. Roy. Belg.},
  volume  = {18},
  pages   = {1},
  year    = {1940}
}

@misc{Cao,
  author    = {Cao, Zheng},
  date  = {2026},
  howpublished = "Private communication"}

@article{Hatta:2014gga,
    author = "Hatta, Yoshitaka and Noronha, Jorge and Xiao, Bo-Wen",
    title = "{A systematic study of exact solutions in second-order conformal hydrodynamics}",
    eprint = "1403.7693",
    archivePrefix = "arXiv",
    primaryClass = "hep-th",
    reportNumber = "YITP-14-21",
    doi = "10.1103/PhysRevD.89.114011",
    journal = "Phys. Rev. D",
    volume = "89",
    number = "11",
    pages = "114011",
    year = "2014"
}

@book{Weinberg:1972kfs,
    author = "Weinberg, Steven",
    title = "{Gravitation and Cosmology}: {Principles and Applications of the General Theory of Relativity}",
    isbn = "978-0-471-92567-5, 978-0-471-92567-5",
    publisher = "John Wiley and Sons",
    address = "New York",
    year = "1972"
}

@article{Csorgo:2003rt,
    author = {Cs{\"o}rg{\H{o}}, T. and Grassi, F. and Hama, Y. and Kodama, T.},
    title = "{Simple solutions of relativistic hydrodynamics for longitudinally and cylindrically expanding systems}",
    eprint = "nucl-th/0305059",
    archivePrefix = "arXiv",
    doi = "10.1016/S0370-2693(03)00747-0",
    journal = "Phys. Lett. B",
    volume = "565",
    pages = "107--115",
    year = "2003"
}

@Article{Hatta:2014jva,
  author        = {Hatta, Yoshitaka and Noronha, Jorge and Torrieri, Giorgio and Xiao, Bo-Wen},
  journal       = {Phys. Rev. D},
  title         = {{Flow harmonics within an analytically solvable viscous hydrodynamic model}},
  year          = {2014},
  number        = {7},
  pages         = {074026},
  volume        = {90},
  archiveprefix = {arXiv},
  doi           = {10.1103/PhysRevD.90.074026},
  eprint        = {1407.5952},
  primaryclass  = {hep-ph},
  reportnumber  = {YITP-14-59},
}

@Article{Gubser:2010ze,
  author        = {Gubser, Steven S.},
  journal       = {Phys. Rev. D},
  title         = {{Symmetry constraints on generalizations of Bjorken flow}},
  year          = {2010},
  pages         = {085027},
  volume        = {82},
  archiveprefix = {arXiv},
  doi           = {10.1103/PhysRevD.82.085027},
  eprint        = {1006.0006},
  primaryclass  = {hep-th},
  reportnumber  = {PUPT-2340},
}

@Article{Gubser:2010ui,
  author        = {Gubser, Steven S. and Yarom, Amos},
  journal       = {Nucl. Phys. B},
  title         = {{Conformal hydrodynamics in Minkowski and de Sitter spacetimes}},
  year          = {2011},
  pages         = {469--511},
  volume        = {846},
  archiveprefix = {arXiv},
  doi           = {10.1016/j.nuclphysb.2011.01.012},
  eprint        = {1012.1314},
  primaryclass  = {hep-th},
  reportnumber  = {PUPT-2358},
}

@Article{Becattini:2021suc,
  author        = {Becattini, F. and Buzzegoli, M. and Palermo, A.},
  journal       = {Phys. Lett. B},
  title         = {{Spin-thermal shear coupling in a relativistic fluid}},
  year          = {2021},
  pages         = {136519},
  volume        = {820},
  archiveprefix = {arXiv},
  doi           = {10.1016/j.physletb.2021.136519},
  eprint        = {2103.10917},
  primaryclass  = {nucl-th},
}

@Article{Becattini:2021iol,
  author        = {Becattini, F. and Buzzegoli, M. and Inghirami, G. and Karpenko, I. and Palermo, A.},
  journal       = {Phys. Rev. Lett.},
  title         = {{Local Polarization and Isothermal Local Equilibrium in Relativistic Heavy Ion Collisions}},
  year          = {2021},
  number        = {27},
  pages         = {272302},
  volume        = {127},
  archiveprefix = {arXiv},
  doi           = {10.1103/PhysRevLett.127.272302},
  eprint        = {2103.14621},
  primaryclass  = {nucl-th},
}

@Article{Liu:2021uhn,
  author        = {Liu, Shuai Y. F. and Yin, Yi},
  journal       = {JHEP\,},
  title         = {{Spin polarization induced by the hydrodynamic gradients}},
  year          = {2021},
  pages         = {188},
  volume        = {07},
  archiveprefix = {arXiv},
  doi           = {10.1007/JHEP07(2021)188},
  eprint        = {2103.09200},
  primaryclass  = {hep-ph},
}

@Article{Fu:2021pok,
  author        = {Fu, Baochi and Liu, Shuai Y. F. and Pang, Longgang and Song, Huichao and Yin, Yi},
  journal       = {Phys. Rev. Lett.},
  title         = {{Shear-Induced Spin Polarization in Heavy-Ion Collisions}},
  year          = {2021},
  number        = {14},
  pages         = {142301},
  volume        = {127},
  archiveprefix = {arXiv},
  doi           = {10.1103/PhysRevLett.127.142301},
  eprint        = {2103.10403},
  primaryclass  = {hep-ph},
}

@Unpublished{Sheng:2025cjk,
  author        = {Sheng, Xin-Li and Becattini, Francesco and Roselli, Daniele},
  title         = {{An improved formula for Wigner function and spin polarization in a decoupling relativistic fluid at local thermodynamic equilibrium}},
  month         = {9},
  year          = {2025},
  archiveprefix = {arXiv},
  eprint        = {2509.14301},
  primaryclass  = {nucl-th},
}

@Article{Yi:2024kwu,
  author        = {Yi, Cong and Wu, Xiang-Yu and Zhu, Jie and Pu, Shi and Qin, Guang-You},
  journal       = {Phys. Rev. C},
  title         = {{Spin polarization of {\ensuremath{\Lambda}} hyperons along the beam direction in p+Pb collisions at sNN=8.16 TeV using hydrodynamic approaches}},
  year          = {2025},
  number        = {4},
  pages         = {044901},
  volume        = {111},
  archiveprefix = {arXiv},
  doi           = {10.1103/PhysRevC.111.044901},
  eprint        = {2408.04296},
  primaryclass  = {hep-ph},
}

@Article{STAR:2019erd,
  author        = {Adam, Jaroslav and others},
  journal       = {Phys. Rev. Lett.},
  title         = {{Polarization of $\Lambda$ ($\bar{\Lambda}$) hyperons along the beam direction in Au+Au collisions at $\sqrt{s_{_{NN}}}$ = 200 GeV}},
  year          = {2019},
  number        = {13},
  pages         = {132301},
  volume        = {123},
  archiveprefix = {arXiv},
  collaboration = {STAR},
  doi           = {10.1103/PhysRevLett.123.132301},
  eprint        = {1905.11917},
  primaryclass  = {nucl-ex},
}

@Article{Bjorken:1982qr,
  author       = {Bjorken, J. D.},
  journal      = {Phys. Rev. D},
  title        = {{Highly Relativistic Nucleus-Nucleus Collisions: The Central Rapidity Region}},
  year         = {1983},
  pages        = {140--151},
  volume       = {27},
  doi          = {10.1103/PhysRevD.27.140},
  reportnumber = {FERMILAB-PUB-82-044-THY, FERMILAB-PUB-82-044-T},
}

@Book{Henner2009,
  author    = {Victor Henner and Tatyana Belozerova and Kyle Forinash},
  publisher = {A K Peters/CRC Press},
  title     = {Mathematical Methods in Physics: Partial Differential Equations, Fourier Series, and Special Functions},
  year      = {2009},
  address   = {New York},
  isbn      = {978-0-429-06252-0},
  doi       = {10.1201/b10695},
}

@Article{STAR:2023eck,
  author        = {Abdulhamid, Muhammad and others},
  journal       = {Phys. Rev. Lett.},
  title         = {{Hyperon Polarization along the Beam Direction Relative to the Second and Third Harmonic Event Planes in Isobar Collisions at sNN=200{\,}{\,}GeV}},
  year          = {2023},
  number        = {20},
  pages         = {202301},
  volume        = {131},
  archiveprefix = {arXiv},
  collaboration = {STAR},
  doi           = {10.1103/PhysRevLett.131.202301},
  eprint        = {2303.09074},
  primaryclass  = {nucl-ex},
}

@Article{Liang:2004ph,
  author        = {Liang, Zuo-Tang and Wang, Xin-Nian},
  journal       = {Phys. Rev. Lett.},
  title         = {{Globally polarized quark-gluon plasma in non-central A+A collisions}},
  year          = {2005},
  note          = {[Erratum: Phys.Rev.Lett. 96, 039901 (2006)]},
  pages         = {102301},
  volume        = {94},
  archiveprefix = {arXiv},
  doi           = {10.1103/PhysRevLett.94.102301},
  eprint        = {nucl-th/0410079},
  reportnumber  = {LBNL-56383},
}

@Unpublished{Voloshin:2004ha,
  author        = {Voloshin, Sergei A.},
  title         = {{Polarized secondary particles in unpolarized high energy hadron-hadron collisions?}},
  month         = {10},
  year          = {2004},
  archiveprefix = {arXiv},
  eprint        = {nucl-th/0410089},
}

@Article{Gao:2007bc,
  author        = {Gao, Jian-Hua and Chen, Shou-Wan and Deng, Wei-tian and Liang, Zuo-Tang and Wang, Qun and Wang, Xin-Nian},
  journal       = {Phys. Rev. C},
  title         = {{Global quark polarization in non-central A+A collisions}},
  year          = {2008},
  pages         = {044902},
  volume        = {77},
  archiveprefix = {arXiv},
  doi           = {10.1103/PhysRevC.77.044902},
  eprint        = {0710.2943},
  primaryclass  = {nucl-th},
  reportnumber  = {LBNL-63515},
}

@Article{STAR:2017ckg,
  author        = {Adamczyk, L. and others},
  journal       = {Nature},
  title         = {{Global $\Lambda$ hyperon polarization in nuclear collisions: evidence for the most vortical fluid}},
  year          = {2017},
  pages         = {62--65},
  volume        = {548},
  archiveprefix = {arXiv},
  collaboration = {STAR},
  doi           = {10.1038/nature23004},
  eprint        = {1701.06657},
  primaryclass  = {nucl-ex},
}

@Article{HADES:2022enx,
  author        = {Abou Yassine, R. and others},
  journal       = {Phys. Lett. B},
  title         = {{Measurement of global polarization of {\ensuremath{\Lambda}} hyperons in few-GeV heavy-ion collisions}},
  year          = {2022},
  pages         = {137506},
  volume        = {835},
  archiveprefix = {arXiv},
  collaboration = {HADES},
  doi           = {10.1016/j.physletb.2022.137506},
  eprint        = {2207.05160},
  primaryclass  = {nucl-ex},
}

@Article{ALICE:2019onw,
  author        = {Acharya, Shreyasi and others},
  journal       = {Phys. Rev. C},
  title         = {{Global polarization of $\Lambda \bar \Lambda$ hyperons in Pb-Pb collisions at $\sqrt {s_{NN}}$ = 2.76 and 5.02 TeV}},
  year          = {2020},
  note          = {[Erratum: Phys.Rev.C 105, 029902 (2022)]},
  number        = {4},
  pages         = {044611},
  volume        = {101},
  archiveprefix = {arXiv},
  collaboration = {ALICE},
  doi           = {10.1103/PhysRevC.101.044611},
  eprint        = {1909.01281},
  primaryclass  = {nucl-ex},
  reportnumber  = {CERN-EP-2019-173},
}

@Article{Becattini:2013fla,
  author        = {Becattini, F. and Chandra, V. and Del Zanna, L. and Grossi, E.},
  journal       = {Annals Phys.},
  title         = {{Relativistic distribution function for particles with spin at local thermodynamical equilibrium}},
  year          = {2013},
  pages         = {32--49},
  volume        = {338},
  archiveprefix = {arXiv},
  doi           = {10.1016/j.aop.2013.07.004},
  eprint        = {1303.3431},
  primaryclass  = {nucl-th},
}

@Article{Fang:2016vpj,
  author        = {Fang, Ren-hong and Pang, Long-gang and Wang, Qun and Wang, Xin-nian},
  journal       = {Phys. Rev. C},
  title         = {{Polarization of massive fermions in a vortical fluid}},
  year          = {2016},
  number        = {2},
  pages         = {024904},
  volume        = {94},
  archiveprefix = {arXiv},
  doi           = {10.1103/PhysRevC.94.024904},
  eprint        = {1604.04036},
  primaryclass  = {nucl-th},
  reportnumber  = {ICTS-USTC-16-05},
}

@Article{Xia:2018tes,
  author        = {Xia, Xiao-Liang and Li, Hui and Tang, Ze-Bo and Wang, Qun},
  journal       = {Phys. Rev. C},
  title         = {{Probing vorticity structure in heavy-ion collisions by local $\Lambda$ polarization}},
  year          = {2018},
  pages         = {024905},
  volume        = {98},
  archiveprefix = {arXiv},
  doi           = {10.1103/PhysRevC.98.024905},
  eprint        = {1803.00867},
  primaryclass  = {nucl-th},
}

@Article{Karpenko:2016jyx,
  author        = {Karpenko, I. and Becattini, F.},
  journal       = {Eur. Phys. J. C},
  title         = {{Study of $\Lambda $ polarization in relativistic nuclear collisions at $\sqrt{s_\mathrm {NN}}=7.7$ {\textendash}200 GeV}},
  year          = {2017},
  number        = {4},
  pages         = {213},
  volume        = {77},
  archiveprefix = {arXiv},
  doi           = {10.1140/epjc/s10052-017-4765-1},
  eprint        = {1610.04717},
  primaryclass  = {nucl-th},
}

@Article{Sun:2017xhx,
  author        = {Sun, Yifeng and Ko, Che Ming},
  journal       = {Phys. Rev. C},
  title         = {{$\Lambda$ hyperon polarization in relativistic heavy ion collisions from a chiral kinetic approach}},
  year          = {2017},
  number        = {2},
  pages         = {024906},
  volume        = {96},
  archiveprefix = {arXiv},
  doi           = {10.1103/PhysRevC.96.024906},
  eprint        = {1706.09467},
  primaryclass  = {nucl-th},
}

@Article{Li:2017slc,
  author        = {Li, Hui and Pang, Long-Gang and Wang, Qun and Xia, Xiao-Liang},
  journal       = {Phys. Rev. C},
  title         = {{Global $\Lambda$ polarization in heavy-ion collisions from a transport model}},
  year          = {2017},
  number        = {5},
  pages         = {054908},
  volume        = {96},
  archiveprefix = {arXiv},
  doi           = {10.1103/PhysRevC.96.054908},
  eprint        = {1704.01507},
  primaryclass  = {nucl-th},
}

@Article{Wei:2018zfb,
  author        = {Wei, De-Xian and Deng, Wei-Tian and Huang, Xu-Guang},
  journal       = {Phys. Rev. C},
  title         = {{Thermal vorticity and spin polarization in heavy-ion collisions}},
  year          = {2019},
  number        = {1},
  pages         = {014905},
  volume        = {99},
  archiveprefix = {arXiv},
  doi           = {10.1103/PhysRevC.99.014905},
  eprint        = {1810.00151},
  primaryclass  = {nucl-th},
}

@Article{Fu:2020oxj,
  author        = {Fu, Baochi and Xu, Kai and Huang, Xu-Guang and Song, Huichao},
  journal       = {Phys. Rev. C},
  title         = {{Hydrodynamic study of hyperon spin polarization in relativistic heavy ion collisions}},
  year          = {2021},
  number        = {2},
  pages         = {024903},
  volume        = {103},
  archiveprefix = {arXiv},
  doi           = {10.1103/PhysRevC.103.024903},
  eprint        = {2011.03740},
  primaryclass  = {nucl-th},
}

@Article{Lei:2021mvp,
  author        = {Lei, Anke and Wang, Dujuan and Zhou, Dai-Mei and Sa, Ben-Hao and Csernai, Laszlo Pal},
  journal       = {Phys. Rev. C},
  title         = {{Vorticity and $\Lambda$ polarization in the microscopic transport model PACIAE}},
  year          = {2021},
  number        = {5},
  pages         = {054903},
  volume        = {104},
  archiveprefix = {arXiv},
  doi           = {10.1103/PhysRevC.104.054903},
  eprint        = {2110.13485},
  primaryclass  = {nucl-th},
}

@Article{Vitiuk:2019rfv,
  author        = {Vitiuk, O. and Bravina, L. V. and Zabrodin, E. E.},
  journal       = {Phys. Lett. B},
  title         = {{Is different $\Lambda$ and $\bar \Lambda$ polarization caused by different spatio-temporal freeze-out picture?}},
  year          = {2020},
  pages         = {135298},
  volume        = {803},
  archiveprefix = {arXiv},
  doi           = {10.1016/j.physletb.2020.135298},
  eprint        = {1910.06292},
  primaryclass  = {hep-ph},
}

@Article{Becattini:2017gcx,
  author        = {Becattini, F. and Karpenko, Iu.},
  journal       = {Phys. Rev. Lett.},
  title         = {{Collective Longitudinal Polarization in Relativistic Heavy-Ion Collisions at Very High Energy}},
  year          = {2018},
  number        = {1},
  pages         = {012302},
  volume        = {120},
  archiveprefix = {arXiv},
  doi           = {10.1103/PhysRevLett.120.012302},
  eprint        = {1707.07984},
  primaryclass  = {nucl-th},
}

@Article{Florkowski:2019voj,
  author        = {Florkowski, Wojciech and Kumar, Avdhesh and Ryblewski, Radoslaw and Mazeliauskas, Aleksas},
  journal       = {Phys. Rev. C},
  title         = {{Longitudinal spin polarization in a thermal model}},
  year          = {2019},
  number        = {5},
  pages         = {054907},
  volume        = {100},
  archiveprefix = {arXiv},
  doi           = {10.1103/PhysRevC.100.054907},
  eprint        = {1904.00002},
  primaryclass  = {nucl-th},
}

@Article{Wu:2019eyi,
  author        = {Wu, Hong-Zhong and Pang, Long-Gang and Huang, Xu-Guang and Wang, Qun},
  journal       = {Phys. Rev. Research.},
  title         = {{Local spin polarization in high energy heavy ion collisions}},
  year          = {2019},
  pages         = {033058},
  volume        = {1},
  archiveprefix = {arXiv},
  doi           = {10.1103/PhysRevResearch.1.033058},
  eprint        = {1906.09385},
  primaryclass  = {nucl-th},
}

@Article{Becattini:2019ntv,
  author        = {Becattini, Francesco and Cao, Gaoqing and Speranza, Enrico},
  journal       = {Eur. Phys. J. C},
  title         = {{Polarization transfer in hyperon decays and its effect in relativistic nuclear collisions}},
  year          = {2019},
  number        = {9},
  pages         = {741},
  volume        = {79},
  archiveprefix = {arXiv},
  doi           = {10.1140/epjc/s10052-019-7213-6},
  eprint        = {1905.03123},
  primaryclass  = {nucl-th},
}

@Article{ALICE:2021pzu,
  author        = {Acharya, Shreyasi and others},
  journal       = {Phys. Rev. Lett.},
  title         = {{Polarization of $\Lambda$ and $\bar \Lambda$ Hyperons along the Beam Direction in Pb-Pb Collisions at $\sqrt {s_{NN}}$=5.02{\,}{\,}TeV}},
  year          = {2022},
  number        = {17},
  pages         = {172005},
  volume        = {128},
  archiveprefix = {arXiv},
  collaboration = {ALICE},
  doi           = {10.1103/PhysRevLett.128.172005},
  eprint        = {2107.11183},
  primaryclass  = {nucl-ex},
  reportnumber  = {CERN-EP-2021-148},
}

@Article{CMS:2025nqr,
  author        = {Hayrapetyan, Aram and others},
  journal       = {Phys. Rev. Lett.},
  title         = {{Observation of {\ensuremath{\Lambda}} Hyperon Local Polarization in p-Pb Collisions at sNN=8.16{\,}{\,}TeV}},
  year          = {2025},
  number        = {13},
  pages         = {132301},
  volume        = {135},
  archiveprefix = {arXiv},
  collaboration = {CMS},
  doi           = {10.1103/6ywq-gm61},
  eprint        = {2502.07898},
  primaryclass  = {nucl-ex},
  reportnumber  = {CMS-HIN-24-002, CERN-EP-2024-328},
}

@Article{Xia:2019fjf,
  author        = {Xia, Xiao-Liang and Li, Hui and Huang, Xu-Guang and Huang, Huan Zhong},
  journal       = {Phys. Rev. C},
  title         = {{Feed-down effect on {\ensuremath{\Lambda}} spin polarization}},
  year          = {2019},
  number        = {1},
  pages         = {014913},
  volume        = {100},
  archiveprefix = {arXiv},
  doi           = {10.1103/PhysRevC.100.014913},
  eprint        = {1905.03120},
  primaryclass  = {nucl-th},
}

@Article{Gao:2012ix,
  author        = {Gao, Jian-Hua and Liang, Zuo-Tang and Pu, Shi and Wang, Qun and Wang, Xin-Nian},
  journal       = {Phys. Rev. Lett.},
  title         = {{Chiral Anomaly and Local Polarization Effect from Quantum Kinetic Approach}},
  year          = {2012},
  pages         = {232301},
  volume        = {109},
  archiveprefix = {arXiv},
  doi           = {10.1103/PhysRevLett.109.232301},
  eprint        = {1203.0725},
  primaryclass  = {hep-ph},
  reportnumber  = {USTC-ICTS-12-02},
}

@Article{Chen:2012ca,
  author        = {Chen, Jiunn-Wei and Pu, Shi and Wang, Qun and Wang, Xin-Nian},
  journal       = {Phys. Rev. Lett.},
  title         = {{Berry Curvature and Four-Dimensional Monopoles in the Relativistic Chiral Kinetic Equation}},
  year          = {2013},
  number        = {26},
  pages         = {262301},
  volume        = {110},
  archiveprefix = {arXiv},
  doi           = {10.1103/PhysRevLett.110.262301},
  eprint        = {1210.8312},
  primaryclass  = {hep-th},
  reportnumber  = {USTC-ICTS-12-14},
}

@Article{Hidaka:2016yjf,
  author        = {Hidaka, Yoshimasa and Pu, Shi and Yang, Di-Lun},
  journal       = {Phys. Rev. D},
  title         = {{Relativistic Chiral Kinetic Theory from Quantum Field Theories}},
  year          = {2017},
  number        = {9},
  pages         = {091901},
  volume        = {95},
  archiveprefix = {arXiv},
  doi           = {10.1103/PhysRevD.95.091901},
  eprint        = {1612.04630},
  primaryclass  = {hep-th},
}

@Article{Hidaka:2017auj,
  author        = {Hidaka, Yoshimasa and Pu, Shi and Yang, Di-Lun},
  journal       = {Phys. Rev. D},
  title         = {{Nonlinear Responses of Chiral Fluids from Kinetic Theory}},
  year          = {2018},
  number        = {1},
  pages         = {016004},
  volume        = {97},
  archiveprefix = {arXiv},
  doi           = {10.1103/PhysRevD.97.016004},
  eprint        = {1710.00278},
  primaryclass  = {hep-th},
  reportnumber  = {RIKEN-QHP-260, RIKEN-iTHEMS-Report-17},
}

@Article{Weickgenannt:2019dks,
  author        = {Weickgenannt, Nora and Sheng, Xin-Li and Speranza, Enrico and Wang, Qun and Rischke, Dirk H.},
  journal       = {Phys. Rev. D},
  title         = {{Kinetic theory for massive spin-1/2 particles from the Wigner-function formalism}},
  year          = {2019},
  number        = {5},
  pages         = {056018},
  volume        = {100},
  archiveprefix = {arXiv},
  doi           = {10.1103/PhysRevD.100.056018},
  eprint        = {1902.06513},
  primaryclass  = {hep-ph},
}

@Article{Weickgenannt:2020aaf,
  author        = {Weickgenannt, Nora and Speranza, Enrico and Sheng, Xin-li and Wang, Qun and Rischke, Dirk H.},
  journal       = {Phys. Rev. Lett.},
  title         = {{Generating Spin Polarization from Vorticity through Nonlocal Collisions}},
  year          = {2021},
  number        = {5},
  pages         = {052301},
  volume        = {127},
  archiveprefix = {arXiv},
  doi           = {10.1103/PhysRevLett.127.052301},
  eprint        = {2005.01506},
  primaryclass  = {hep-ph},
}

@Article{Gao:2019znl,
  author        = {Gao, Jian-Hua and Liang, Zuo-Tang},
  journal       = {Phys. Rev. D},
  title         = {{Relativistic Quantum Kinetic Theory for Massive Fermions and Spin Effects}},
  year          = {2019},
  number        = {5},
  pages         = {056021},
  volume        = {100},
  archiveprefix = {arXiv},
  doi           = {10.1103/PhysRevD.100.056021},
  eprint        = {1902.06510},
  primaryclass  = {hep-ph},
}

@Article{Liu:2020flb,
  author        = {Liu, Yu-Chen and Mameda, Kazuya and Huang, Xu-Guang},
  journal       = {Chin. Phys. C},
  title         = {{Covariant Spin Kinetic Theory I: Collisionless Limit}},
  year          = {2020},
  note          = {[Erratum: Chin.Phys.C 45, 089001 (2021)]},
  number        = {9},
  pages         = {094101},
  volume        = {44},
  archiveprefix = {arXiv},
  doi           = {10.1088/1674-1137/ac009b},
  eprint        = {2002.03753},
  primaryclass  = {hep-ph},
}

@Article{Weickgenannt:2021cuo,
  author        = {Weickgenannt, Nora and Speranza, Enrico and Sheng, Xin-li and Wang, Qun and Rischke, Dirk H.},
  journal       = {Phys. Rev. D},
  title         = {{Derivation of the nonlocal collision term in the relativistic Boltzmann equation for massive spin-1/2 particles from quantum field theory}},
  year          = {2021},
  number        = {1},
  pages         = {016022},
  volume        = {104},
  archiveprefix = {arXiv},
  doi           = {10.1103/PhysRevD.104.016022},
  eprint        = {2103.04896},
  primaryclass  = {nucl-th},
}

@Article{Sheng:2021kfc,
  author        = {Sheng, Xin-Li and Weickgenannt, Nora and Speranza, Enrico and Rischke, Dirk H. and Wang, Qun},
  journal       = {Phys. Rev. D},
  title         = {{From Kadanoff-Baym to Boltzmann equations for massive spin-1/2 fermions}},
  year          = {2021},
  number        = {1},
  pages         = {016029},
  volume        = {104},
  archiveprefix = {arXiv},
  doi           = {10.1103/PhysRevD.104.016029},
  eprint        = {2103.10636},
  primaryclass  = {nucl-th},
  reportnumber  = {USTC-ICTS/PCFT-21-12},
}

@Article{Fang:2022ttm,
  author        = {Fang, Shuo and Pu, Shi and Yang, Di-Lun},
  journal       = {Phys. Rev. D},
  title         = {{Quantum kinetic theory for dynamical spin polarization from QED-type interaction}},
  year          = {2022},
  number        = {1},
  pages         = {016002},
  volume        = {106},
  archiveprefix = {arXiv},
  doi           = {10.1103/PhysRevD.106.016002},
  eprint        = {2204.11519},
  primaryclass  = {hep-ph},
}

@Article{Hidaka:2022dmn,
  author        = {Hidaka, Yoshimasa and Pu, Shi and Wang, Qun and Yang, Di-Lun},
  journal       = {Prog. Part. Nucl. Phys.},
  title         = {{Foundations and applications of quantum kinetic theory}},
  year          = {2022},
  pages         = {103989},
  volume        = {127},
  archiveprefix = {arXiv},
  doi           = {10.1016/j.ppnp.2022.103989},
  eprint        = {2201.07644},
  primaryclass  = {hep-ph},
  reportnumber  = {KEK-TH-2390, J-PARC-TH-0267, RIKEN-iTHEMS-Report-22},
}

@Article{Fang:2023bbw,
  author        = {Fang, Shuo and Pu, Shi and Yang, Di-Lun},
  journal       = {Phys. Rev. D},
  title         = {{Spin polarization and spin alignment from quantum kinetic theory with self-energy corrections}},
  year          = {2024},
  number        = {3},
  pages         = {034034},
  volume        = {109},
  archiveprefix = {arXiv},
  doi           = {10.1103/PhysRevD.109.034034},
  eprint        = {2311.15197},
  primaryclass  = {hep-ph},
}

@Article{Fang:2024vds,
  author        = {Fang, Shuo and Pu, Shi},
  journal       = {Phys. Rev. D},
  title         = {{Collisional corrections to spin polarization from quantum kinetic theory using Chapman-Enskog expansion}},
  year          = {2025},
  number        = {3},
  pages         = {034015},
  volume        = {111},
  archiveprefix = {arXiv},
  doi           = {10.1103/PhysRevD.111.034015},
  eprint        = {2408.09877},
  primaryclass  = {hep-ph},
}

@Article{Florkowski:2024bfw,
  author        = {Florkowski, Wojciech and Hontarenko, Mykhailo},
  journal       = {Phys. Rev. Lett.},
  title         = {{Generalized Thermodynamic Relations for Perfect Spin Hydrodynamics}},
  year          = {2025},
  number        = {8},
  pages         = {082302},
  volume        = {134},
  archiveprefix = {arXiv},
  doi           = {10.1103/PhysRevLett.134.082302},
  eprint        = {2405.03263},
  primaryclass  = {hep-ph},
}

@Article{Bhadury:2022ulr,
  author        = {Bhadury, Samapan and Florkowski, Wojciech and Jaiswal, Amaresh and Kumar, Avdhesh and Ryblewski, Radoslaw},
  journal       = {Phys. Rev. Lett.},
  title         = {{Relativistic Spin Magnetohydrodynamics}},
  year          = {2022},
  number        = {19},
  pages         = {192301},
  volume        = {129},
  archiveprefix = {arXiv},
  doi           = {10.1103/PhysRevLett.129.192301},
  eprint        = {2204.01357},
  primaryclass  = {nucl-th},
}

@Article{Florkowski:2018fap,
  author        = {Florkowski, Wojciech and Kumar, Avdhesh and Ryblewski, Radoslaw},
  journal       = {Prog. Part. Nucl. Phys.},
  title         = {{Relativistic hydrodynamics for spin-polarized fluids}},
  year          = {2019},
  pages         = {103709},
  volume        = {108},
  archiveprefix = {arXiv},
  doi           = {10.1016/j.ppnp.2019.07.001},
  eprint        = {1811.04409},
  primaryclass  = {nucl-th},
}

@Article{Becattini:2018duy,
  author        = {Becattini, F. and Florkowski, Wojciech and Speranza, Enrico},
  journal       = {Phys. Lett. B},
  title         = {{Spin tensor and its role in non-equilibrium thermodynamics}},
  year          = {2019},
  pages         = {419--425},
  volume        = {789},
  archiveprefix = {arXiv},
  doi           = {10.1016/j.physletb.2018.12.016},
  eprint        = {1807.10994},
  primaryclass  = {hep-th},
}

@Article{Florkowski:2017ruc,
  author        = {Florkowski, Wojciech and Friman, Bengt and Jaiswal, Amaresh and Speranza, Enrico},
  journal       = {Phys. Rev. C},
  title         = {{Relativistic fluid dynamics with spin}},
  year          = {2018},
  number        = {4},
  pages         = {041901},
  volume        = {97},
  archiveprefix = {arXiv},
  doi           = {10.1103/PhysRevC.97.041901},
  eprint        = {1705.00587},
  primaryclass  = {nucl-th},
}

@Article{Cao:2022aku,
  author        = {Cao, Zheng and Hattori, Koichi and Hongo, Masaru and Huang, Xu-Guang and Taya, Hidetoshi},
  journal       = {PTEP},
  title         = {{Gyrohydrodynamics: Relativistic spinful fluid with strong vorticity}},
  year          = {2022},
  number        = {7},
  pages         = {071D01},
  volume        = {2022},
  archiveprefix = {arXiv},
  doi           = {10.1093/ptep/ptac091},
  eprint        = {2205.08051},
  primaryclass  = {hep-th},
  reportnumber  = {RIKEN-iTHEMS-Report-22},
}

@Article{Hattori:2019lfp,
  author        = {Hattori, Koichi and Hongo, Masaru and Huang, Xu-Guang and Matsuo, Mamoru and Taya, Hidetoshi},
  journal       = {Phys. Lett. B},
  title         = {{Fate of spin polarization in a relativistic fluid: An entropy-current analysis}},
  year          = {2019},
  pages         = {100--106},
  volume        = {795},
  archiveprefix = {arXiv},
  doi           = {10.1016/j.physletb.2019.05.040},
  eprint        = {1901.06615},
  primaryclass  = {hep-th},
  reportnumber  = {RIKEN-iTHEMS-Report-19, YITP-19-15},
}

@Article{Hongo:2021ona,
  author        = {Hongo, Masaru and Huang, Xu-Guang and Kaminski, Matthias and Stephanov, Mikhail and Yee, Ho-Ung},
  journal       = {JHEP\,},
  title         = {{Relativistic spin hydrodynamics with torsion and linear response theory for spin relaxation}},
  year          = {2021},
  pages         = {150},
  volume        = {11},
  archiveprefix = {arXiv},
  doi           = {10.1007/JHEP11(2021)150},
  eprint        = {2107.14231},
  primaryclass  = {hep-th},
  reportnumber  = {RIKEN-iTHEMS-Report-21},
}

@Article{Li:2020eon,
  author        = {Li, Shiyong and Stephanov, Mikhail A. and Yee, Ho-Ung},
  journal       = {Phys. Rev. Lett.},
  title         = {{Nondissipative Second-Order Transport, Spin, and Pseudogauge Transformations in Hydrodynamics}},
  year          = {2021},
  number        = {8},
  pages         = {082302},
  volume        = {127},
  archiveprefix = {arXiv},
  doi           = {10.1103/PhysRevLett.127.082302},
  eprint        = {2011.12318},
  primaryclass  = {hep-th},
}

@Article{Montenegro:2017lvf,
  author        = {Montenegro, David and Tinti, Leonardo and Torrieri, Giorgio},
  journal       = {Phys. Rev. D},
  title         = {{Sound waves and vortices in a polarized relativistic fluid}},
  year          = {2017},
  number        = {7},
  pages         = {076016},
  volume        = {96},
  archiveprefix = {arXiv},
  doi           = {10.1103/PhysRevD.96.076016},
  eprint        = {1703.03079},
  primaryclass  = {hep-th},
}

@Article{Weickgenannt:2022qvh,
  author        = {Weickgenannt, Nora and Wagner, David and Speranza, Enrico and Rischke, Dirk H.},
  journal       = {Phys. Rev. D},
  title         = {{Relativistic dissipative spin hydrodynamics from kinetic theory with a nonlocal collision term}},
  year          = {2022},
  number        = {9},
  pages         = {L091901},
  volume        = {106},
  archiveprefix = {arXiv},
  doi           = {10.1103/PhysRevD.106.L091901},
  eprint        = {2208.01955},
  primaryclass  = {nucl-th},
}

@Article{Weickgenannt:2022zxs,
  author        = {Weickgenannt, Nora and Wagner, David and Speranza, Enrico and Rischke, Dirk H.},
  journal       = {Phys. Rev. D},
  title         = {{Relativistic second-order dissipative spin hydrodynamics from the method of moments}},
  year          = {2022},
  number        = {9},
  pages         = {096014},
  volume        = {106},
  archiveprefix = {arXiv},
  doi           = {10.1103/PhysRevD.106.096014},
  eprint        = {2203.04766},
  primaryclass  = {nucl-th},
}

@Article{Fukushima:2020ucl,
  author        = {Fukushima, Kenji and Pu, Shi},
  journal       = {Phys. Lett. B},
  title         = {{Spin hydrodynamics and symmetric energy-momentum tensors {\textendash} A current induced by the spin vorticity {\textendash}}},
  year          = {2021},
  pages         = {136346},
  volume        = {817},
  archiveprefix = {arXiv},
  doi           = {10.1016/j.physletb.2021.136346},
  eprint        = {2010.01608},
  primaryclass  = {hep-th},
}

@Unpublished{Fang:2025aig,
  author        = {Fang, Shuo and Fukushima, Kenji and Pu, Shi and Wang, Dong-Lin},
  title         = {{Relativistic spin hydrodynamics with antisymmetric spin tensors and an extension of the Bargmann-Michel-Telegdi equation}},
  month         = {6},
  year          = {2025},
  archiveprefix = {arXiv},
  eprint        = {2506.20698},
  primaryclass  = {nucl-th},
}

@Article{Wang:2021ngp,
  author        = {Wang, Dong-Lin and Fang, Shuo and Pu, Shi},
  journal       = {Phys. Rev. D},
  title         = {{Analytic solutions of relativistic dissipative spin hydrodynamics with Bjorken expansion}},
  year          = {2021},
  number        = {11},
  pages         = {114043},
  volume        = {104},
  archiveprefix = {arXiv},
  doi           = {10.1103/PhysRevD.104.114043},
  eprint        = {2107.11726},
  primaryclass  = {nucl-th},
}

@Article{She:2021lhe,
  author        = {She, Duan and Huang, Anping and Hou, Defu and Liao, Jinfeng},
  journal       = {Sci. Bull.},
  title         = {{Relativistic viscous hydrodynamics with angular momentum}},
  year          = {2022},
  pages         = {2265--2268},
  volume        = {67},
  archiveprefix = {arXiv},
  doi           = {10.1016/j.scib.2022.10.020},
  eprint        = {2105.04060},
  primaryclass  = {nucl-th},
}

@Article{Huang:2024ffg,
  author        = {Huang, Xu-Guang},
  journal       = {Nucl. Sci. Tech.},
  title         = {{An introduction to relativistic spin hydrodynamics}},
  year          = {2025},
  number        = {11},
  pages         = {208},
  volume        = {36},
  archiveprefix = {arXiv},
  doi           = {10.1007/s41365-025-01784-3},
  eprint        = {2411.11753},
  primaryclass  = {nucl-th},
}

@Article{Yi:2021ryh,
  author        = {Yi, Cong and Pu, Shi and Yang, Di-Lun},
  journal       = {Phys. Rev. C},
  title         = {{Reexamination of local spin polarization beyond global equilibrium in relativistic heavy ion collisions}},
  year          = {2021},
  number        = {6},
  pages         = {064901},
  volume        = {104},
  archiveprefix = {arXiv},
  doi           = {10.1103/PhysRevC.104.064901},
  eprint        = {2106.00238},
  primaryclass  = {hep-ph},
}

@Article{Ryu:2021lnx,
  author        = {Ryu, Sangwook and Jupic, Vahidin and Shen, Chun},
  journal       = {Phys. Rev. C},
  title         = {{Probing early-time longitudinal dynamics with the {\ensuremath{\Lambda}} hyperon's spin polarization in relativistic heavy-ion collisions}},
  year          = {2021},
  number        = {5},
  pages         = {054908},
  volume        = {104},
  archiveprefix = {arXiv},
  doi           = {10.1103/PhysRevC.104.054908},
  eprint        = {2106.08125},
  primaryclass  = {nucl-th},
}

@Article{Florkowski:2021xvy,
  author        = {Florkowski, Wojciech and Kumar, Avdhesh and Mazeliauskas, Aleksas and Ryblewski, Radoslaw},
  journal       = {Phys. Rev. C},
  title         = {{Effect of thermal shear on longitudinal spin polarization in a thermal model}},
  year          = {2022},
  number        = {6},
  pages         = {064901},
  volume        = {105},
  archiveprefix = {arXiv},
  doi           = {10.1103/PhysRevC.105.064901},
  eprint        = {2112.02799},
  primaryclass  = {hep-ph},
}

@Article{Buzzegoli:2022fxu,
  author        = {Buzzegoli, M. and Becattini, F. and Inghirami, G. and Karpenko, I. and Palermo, A.},
  journal       = {Acta Phys. Polon. Supp.},
  title         = {{Spin-thermal Shear Coupling in Relativistic Nuclear Collisions}},
  year          = {2023},
  number        = {1},
  pages         = {1--A39},
  volume        = {16},
  archiveprefix = {arXiv},
  doi           = {10.5506/APhysPolBSupp.16.1-A39},
  eprint        = {2208.04449},
  primaryclass  = {nucl-th},
}

@Article{Palermo:2022lvh,
  author        = {Palermo, Andrea and Becattini, Francesco and Buzzegoli, Matteo and Inghirami, Gabriele and Karpenko, Iurii},
  journal       = {EPJ Web Conf.},
  title         = {{Local equilibrium and Lambda polarization in high energy heavy ion collisions}},
  year          = {2023},
  pages         = {01026},
  volume        = {276},
  archiveprefix = {arXiv},
  doi           = {10.1051/epjconf/202327601026},
  eprint        = {2208.09874},
  primaryclass  = {nucl-th},
}

@Article{Wu:2022mkr,
  author        = {Wu, Xiang-Yu and Yi, Cong and Qin, Guang-You and Pu, Shi},
  journal       = {Phys. Rev. C},
  title         = {{Local and global polarization of {\ensuremath{\Lambda}} hyperons across RHIC-BES energies: The roles of spin hall effect, initial condition, and baryon diffusion}},
  year          = {2022},
  number        = {6},
  pages         = {064909},
  volume        = {105},
  archiveprefix = {arXiv},
  doi           = {10.1103/PhysRevC.105.064909},
  eprint        = {2204.02218},
  primaryclass  = {hep-ph},
}

@Article{Palermo:2024tza,
  author        = {Palermo, Andrea and Grossi, Eduardo and Karpenko, Iurii and Becattini, Francesco},
  journal       = {Eur. Phys. J. C},
  title         = {{$\Lambda $ polarization in very high energy heavy ion collisions as a probe of the quark{\textendash}gluon plasma formation and properties}},
  year          = {2024},
  number        = {9},
  pages         = {920},
  volume        = {84},
  archiveprefix = {arXiv},
  doi           = {10.1140/epjc/s10052-024-13229-z},
  eprint        = {2404.14295},
  primaryclass  = {nucl-th},
}

@Article{Becattini:2024uha,
  author        = {Becattini, Francesco and Buzzegoli, Matteo and Niida, Takafumi and Pu, Shi and Tang, Ai-Hong and Wang, Qun},
  journal       = {Int. J. Mod. Phys. E},
  title         = {{Spin polarization in relativistic heavy-ion collisions}},
  year          = {2024},
  number        = {06},
  pages         = {2430006},
  volume        = {33},
  archiveprefix = {arXiv},
  doi           = {10.1142/9789811294679_0005},
  eprint        = {2402.04540},
  primaryclass  = {nucl-th},
}

@Article{Niida:2024ntm,
  author        = {Niida, Takafumi and Voloshin, Sergei A.},
  journal       = {Int. J. Mod. Phys. E},
  title         = {{Polarization phenomenon in heavy-ion collisions}},
  year          = {2024},
  number        = {09},
  pages         = {2430010},
  volume        = {33},
  archiveprefix = {arXiv},
  doi           = {10.1142/S0218301324300108},
  eprint        = {2404.11042},
  primaryclass  = {nucl-ex},
}

@Article{Alzhrani:2022dpi,
  author        = {Alzhrani, Sahr and Ryu, Sangwook and Shen, Chun},
  journal       = {Phys. Rev. C},
  title         = {{{\ensuremath{\Lambda}} spin polarization in event-by-event relativistic heavy-ion collisions}},
  year          = {2022},
  number        = {1},
  pages         = {014905},
  volume        = {106},
  archiveprefix = {arXiv},
  doi           = {10.1103/PhysRevC.106.014905},
  eprint        = {2203.15718},
  primaryclass  = {nucl-th},
}

@Article{Arslan:2024dwi,
  author        = {Arslan, Anum and Dong, Wen-Bo and Ma, Guo-Liang and Pu, Shi and Wang, Qun},
  journal       = {Phys. Rev. C},
  title         = {{Solvable model for spin polarizations with flow-momentum correspondence}},
  year          = {2025},
  number        = {4},
  pages         = {044911},
  volume        = {111},
  archiveprefix = {arXiv},
  doi           = {10.1103/PhysRevC.111.044911},
  eprint        = {2411.17285},
  primaryclass  = {nucl-th},
}

@Unpublished{Arslan:2025tan,
  author        = {Arslan, Anum and Dong, Wen-Bo and Gale, Charles and Jeon, Sangyong and Wang, Qun and Wu, Xiang-Yu},
  title         = {{In-plane transverse polarization in heavy-ion collisions}},
  month         = {8},
  year          = {2025},
  archiveprefix = {arXiv},
  eprint        = {2509.00796},
  primaryclass  = {nucl-th},
}

@Article{Dong:2022yzt,
  author        = {Dong, Wen-Bo and Yin, Yi-Liang and Wang, Qun},
  journal       = {Phys. Rev. C},
  title         = {{Spin Boltzmann equation~for nonrelativistic spin-1/2 fermions}},
  year          = {2022},
  number        = {5},
  pages         = {054909},
  volume        = {106},
  archiveprefix = {arXiv},
  doi           = {10.1103/PhysRevC.106.054909},
  eprint        = {2209.12402},
  primaryclass  = {nucl-th},
}

@Book{Yagi:2005yb,
  author    = {Yagi, K. and Hatsuda, T. and Miake, Y.},
  publisher = {Cambridge University Press},
  title     = {{Quark-gluon plasma: From big bang to little bang}},
  year      = {2005},
  address   = {Cambridge},
  isbn      = {978-0-521-56108-2},
  series    = {Cambridge Monographs on Particle Physics, Nuclear Physics and Cosmology},
  volume    = {23},
}
\bibliographystyle{apsrev4-2}
\end{document}